\documentclass[twocolumn]{emulateapj}
\usepackage[]{natbib}

\shorttitle{Magnetic Fields in Pulsating B Stars}
\shortauthors{Shultz et al.}

\begin{document}

\title{Critical evaluation of magnetic field detections reported for pulsating B-type stars in the light of ESPaDOnS, Narval and reanalyzed FORS1/2 observations\footnotemark[*]\footnotetext[*]{B\lowercase{ased on observations obtained at the }C\lowercase{anada-}F\lowercase{rance-}H\lowercase{awaii }T\lowercase{elescope} (CFHT) \lowercase{which is operated by the }N\lowercase{ational }R\lowercase{esearch }C\lowercase{ouncil of }C\lowercase{anada, the }I\lowercase{nstitut }N\lowercase{ational des }S\lowercase{ciences de l'}U\lowercase{nivers of the }C\lowercase{entre }N\lowercase{ational de la }R\lowercase{echerche }S\lowercase{cientifique of }F\lowercase{rance, and the }U\lowercase{niversity of }H\lowercase{awaii, and on observations obtained using the }N\lowercase{arval spectropolarimeter at the }O\lowercase{bservatoire du }P\lowercase{ic du }M\lowercase{idi (}F\lowercase{rance), which is operated by the }I\lowercase{nstitut }N\lowercase{ational des }S\lowercase{ciences de l'}U\lowercase{nivers }(INSU).}}

\author{M. Shultz, G.A. Wade, J. Grunhut}
\affil{Royal Military College of Canada, PO Box 17000, Station Forces, Kingston, ON K7K 4B4, Canada}
\affil{Department of Physics, Engineering Physics and Astronomy, Queen's University, 99 University Avenue, Kingston, Ontario K7L 3N6, Canada}
        
\author{S. Bagnulo, J.D. Landstreet\altaffilmark{1}}
\affil{Armagh Observatory, College Hill, Armagh, Northern Ireland, UK BT61 9DG}

\author{C. Neiner, E. Alecian}
\affil{LESIA, UMR 8109 du CNRS, Observatoire de Paris, UPMC, Universit\'e Paris Diderot, 5 place Jules Janssen, 92195, Meudon Cedex, France}

\author{D. Hanes}
\affil{Department of Physics, Engineering Physics and Astronomy, Queen's University, 99 University Avenue, Kingston, Ontario K7L 3N6, Canada}
\author{The MiMeS Collaboration}

\altaffiltext{1}{Also the Department of Physics \& Astronomy, University of Western Ontario, London, ON N6A 3K7, Canada}
\begin{abstract}
Recent spectropolarimetric studies of 7 SPB and $\beta$~Cep stars have suggested that photospheric magnetic fields are more common in B-type pulsators than in the general population of B stars, suggesting a significant connection between magnetic and pulsational phenomena. We present an analysis of new and previously published spectropolarimetric observations of these stars. New Stokes $V$ observations obtained with the high-resolution ESPaDOnS and Narval instruments confirm the presence of a magnetic field in one of the stars ($\epsilon$~Lup), but find no evidence of magnetism in 5 others. A re-analysis of the published longitudinal field measurements obtained with the low-resolution FORS1/2 spectropolarimeters finds that the measurements of all stars show more scatter from zero than can be attributed to Gaussian noise, suggesting the presence of a signal and/or systematic under-estimation of error bars. Re-reduction and re-measurement of the FORS1/2 spectra from the ESO archive demonstrates that small changes in reduction procedure lead to substantial changes in the inferred longitudinal field, and substantially reduces the number of field detections at the 3$\sigma$ level. Furthermore, we find that the published periods are not unique solutions to the time series of either the original or the revised FORS1/2 data. We conclude that the reported field detections, proposed periods and field geometry models for $\alpha$ Pyx, 15 CMa, 33 Eri and V1449 Aql are artefacts of the data analysis and reduction procedures, and that magnetic fields at the reported strength are no more common in SPB/$\beta$ Cep stars than in the general population of B stars.
\end{abstract}

\keywords{stars: magnetic --- stars: pulsators --- stars: individual($\alpha$ Pyx, 15 CMa, 33 Eri, $\epsilon$ Lupi, HY Vel, V1449 Aql)}

\section{Introduction}
Magnetism in hot, massive stars appears to be a relatively uncommon phenomenon. Despite tremendous observational resources directed to large-scale surveys since the beginning of the 21st century only a few dozen magnetic OB stars have been firmly identified.  Their rarity notwithstanding, it is apparent that those stars which do host detectable fields have certain common magnetic characteristics: the fields are usually topologically dipolar, with typical inferred strengths of hundreds to thousands of gauss. There often exists a significant obliquity between the magnetic and rotational axes. The measured fields are remarkably stable, persisting over many rotational cycles with no detectable secular change of field strength or geometry. The stars hosting these fields could not, however, be more diverse: they are both old and young, with strong and weak winds, and rotational periods varying from less than a day to decades. Some possess photospheric chemical peculiarities, others winds or circumstellar matter that interacts with the field. Others experience pulsations. 

While magnetic OB stars generally provide fertile ground for exploring important facets of stellar evolution, the pulsating stars are of particular interest. Pulsating stars provide a unique opportunity to study the subtle influence of the magnetic field on changes of stellar structure in near real-time. Pulsating stars also offer the potential to probe their interiors via asteroseismology, yielding information about stellar structure. Pulsating stars have furthermore been at the forefront of recent discussions of the incidence of magnetism in hot stars. Hubrig et al. (2006, 2009) reported magnetic measurements of 61 pulsating or candidate pulsating B-type stars (50 slowly pulsating B (SPB) stars and 19 $\beta$ Cep stars), acquired using the FORS1 spectropolarimeter at the ESO VLT. In this large sample, Hubrig et al. (2009) claimed $3\sigma$ detection of 52 percent of the SPB stars (i.e. 26 of 50 SBP stars)  and 31 percent of the $\beta$ Cep stars (6 of 19 stars). These results suggested that magnetic fields are significantly more common in pulsating B-type stars than in the overall population of B-type stars. However, Silvester et al. (2009), analysing independent magnetic observations of pulsating B-type stars obtained with the ESPaDOnS and Narval spectropolarimeters, found a much lower incidence of magnetic stars. In particular, of 8 stars claimed to be magnetic by Hubrig et al. (2006, 2009) that were common to both the Hubrig et al. and Silvester et al. studies, Silvester et al. (2009) confirmed fields in only 2. Furthermore, the Magnetism in Massive Stars (MiMeS) survey component has found magnetic fields to be no more common in SPB or $\beta$ Cep stars than in B stars in general (Grunhut et al., 2011). These represent serious discrepancies between the results of these two different approaches (investigated in detail by Bagnulo et al., 2012).  

Recently, Hubrig et al. (2011a; hereafter H2011a) reported numerous longitudinal magnetic field measurements of 6 SPB and $\beta$ Cep stars, obtained from low-resolution ($\lambda/\Delta\lambda\sim$2000) spectropolarimetric observations acquired using the FORS1 and FORS2 instruments at ESO's Very Large Telescope (VLT). All six program stars -- 33 Eri (HD 24587, B5V, SPB), $\xi^1$ CMa (HD 46328, B0.7IV, $\beta$~Cep), 15 CMa (HD 50707, B1IV, $\beta$~Cep), $\alpha$ Pyx (HD 74575, B1.5III, $\beta$~Cep),  HY Vel (HD 74560, B3IV, SPB), and $\epsilon$ Lupi (HD 136504, B2IV, $\beta$~Cep) -- were reported by those authors to host photospheric magnetic fields with maximum measured longitudinal components of order 100 G. Models of the surface magnetic field geometry in the context of the Oblique Rotator Model (ORM; including rotational periods $P_{\rm rot}$, rotational axis inclination angles $i$, magnetic obliquity angles $\beta$, and dipolar field strengths $B_{\rm d}$) were reported for 4 of the 6 stars -- 33 Eri, $\xi^1$ CMa, 15 CMa, and $\alpha$ Pyx. The study by H2011a was followed by a similar examination by Hubrig et al. (2011b, hereafter H2011b) of another $\beta$ Cep star, V1449 Aql, based on observations obtained with the SOFIN spectropolarimeter on the Nordic Optical Telescope. H2011b report an apparently coherent sinusoidal modulation of the longitudinal field with a full amplitude of $\sim$1 kG. These results are qualitatively distinguished from most previous studies of magnetism in pulsating stars by the long-term monitoring of a large sample of particular stars. They potentially represent an important contribution to the accumulated knowledge and scientific discussion regarding magnetism of pulsating massive stars.

The aim of the current paper is to further explore the incidence of magnetic fields in pulsating B-type stars by independently investigating and elaborating the results reported by H2011a and H2011b. In Section 2 we summarise the observational data employed in this study: new ESPaDOnS and Narval observations, the published FORS1/2 longitudinal magnetic field measurements of H2011a, and new magnetic field measurements obtained from re-reduction and re-measurement of the FORS1/2 spectra acquired by H2011a (since the SOFIN observations reported by H2011b remain proprietary, we are unable to perform as comprehensive an analysis in this case). In Section 3 we present the magnetic analysis of the ESPaDOnS and Narval spectropolarimetry, the original published FORS1/2 results, and the measurements from re-reduced FORS1/2 observations extracted from the ESO archive. In Section 4 we compare the results between instruments and reductions. In Section 5 we evaluate the models presented by H2011a and H2011b. In Section 6 we discuss the results and their implications for the magnetic fields of the individual targets, the general incidence of magnetic fields in pulsating B-type stars, and the performance and reliability of the FORS1 and FORS2 instruments as stellar magnetometers.

\section{Observations}

\subsection{ESPaDOnS and Narval}


High-resolution (R$\sim$65000) circular polarization (Stokes \textit{V}) spectra were obtained for 6 of the stars studied by H2011a and H2011b using the cross-dispersed \'echelle spectropolarimeters ESPaDOnS (at the Canada-France-Hawaii Telescope) and its clone Narval (at the Telescope Bernardot Lyot (TBL), Pic du Midi Observatory) within the context of the MiMeS Large Program. Altogether, 18 independent observations were obtained between February 2010 and June 2011. Each spectropolarimetric sequence consisted of four individual subexposures taken in different orientations of the half-wave retarders. From each set of four subexposures we derived Stokes $I$ and Stokes $V$ spectra following the double-ratio procedure (e.g. Donati et al., 1997). All frames were processed using the automated reduction package Libre-ESpRIT (Donati et al., 1997), with the Stokes $I$ spectra normalized order-by-order using Legendre polynomials following extraction of Stokes $V$. The peak signal-to-noise ratios (SNRs) per 1.8~km~s$^{-1}$ spectral pixel in the reduced spectra range from about 450 to about 1300. 

As discussed by Semel et al. (1993) and Schnerr et al. (2006), rapid temporal changes in line profile shapes and positions due to pulsation may also introduce spurious signal in the inferred polarisation spectra. In the case of a non-magnetic star, such a phenomenon may result in a false magnetic detection. Spurious signatures may equally be introduced by pulsation when a magnetic field is present, combining with the real Zeeman signature and potentially modifying its shape and amplitude.

Since all targets are either confirmed or candidate $\beta$ Cep or SPB stars, subexposures were timed, as far as possible, to correspond to no more than $\sim$1/80th of a pulsation cycle to ensure that no significant variability occurred during the observation. Of the five observed stars for which periods are known, three ($\epsilon$ Lupi (pulsation periods of 0.155 d (primary) and 0.096 d (secondary); Uytterhoeven et al. 2005, Telting et al. 2006), 33 Eri (0.864 d; de Cat et al., 2005) and HY Vel (3.102 d; Lef\'evre et al., 2009)) were observed with subexposures of hundredths of a period. 15 CMa (0.093 d -- 0.181 d; Shobbrook et al., 2006) was observed with subexposures corresponding to less than 1/80th of the longer period, but about 1/60th of the shorter period. V1449 Aql is a multiperiodic star with a dominant mode period of 0.182 d (Briquet et al., 2009) and several lower-amplitude modes as rapid as 0.119 d. Longer subexposures were required due to its faint apparent magnitude, resulting in subexposures for this star that are somewhat longer than 1/80th of these modes. {$\alpha$ Pyx is a candidate $\beta$ Cep star for which the period is not yet known. However, its subexposure durations were 150 s, and therefore would satisfy the limit described above as long as its pulsation period is longer than 0.14 d.

Given the uncertainties described above, we have performed a number of tests to establish that pulsations did not significantly impact the magnetic diagnosis. We examined the Stokes $I$ profiles associated with each of the subexposures of each observation. For $\epsilon$~Lupi, $\alpha$ Pyx, 33 Eri, 15 CMa and HY Vel, we confirm that no variation of the $I$ profile during an observation is detected above the noise. For V1449 Aql, on the other hand, the $I$ profile is observed to vary during an observation by a maximum of about 10 km~s$^{-1}$ in radial velocity, with a small ($\sim 10\%$) variation in depth, and negligible change in shape. To verify that instrumental instabilities or line profile variations such as those observed in V1449 Aql do not translate into a spurious Stokes $V$ signal, we computed diagnostic null polarization spectra (labeled $N$) by combining the four subexposures in such a way that stellar polarization cancels out (see Donati et al. 1997 or Bagnulo et al. 2009 for more details on the definition of $N$). In no case is any signal detected in $N$ above the level of the noise. Considering the numerical simulations of Schnerr et al. (2006) which demonstrate that spurious signatures due to pulsation appear in both $V$ and $N$, this confirms in the case of V1449 Aql that the observed line variations introduce no spurious signal into Stokes $V$. 

Notwithstanding this conclusion, variability of the profile of V1449 Aql during the observation may result in temporal smearing of the Stokes $I$ and $V$ profiles. To verify that this did not significantly reduce our sensitivity to any magnetic field, we have performed a test in which we generated fictitious non-zero Stokes $V$ profiles by computing the numerical derivative of the Stokes $I$ profile corresponding to each of the subexposures. While these profiles do not correspond to any real magnetic field (e.g. any prediction of the model of H2011b), they have the qualitative characteristics of a real Stokes $V$ Zeeman signature, and in particular reflect the changing shape and position of the line profile of V1449 Aql during the course of a polarimetric observation. They therefore serve to understand the impact of the temporal smearing on Stokes $V$. The fictitious $V$ profiles exhibit small changes from subexposure to subexposure, reflecting the radial velocity, depth and shape variability of Stokes $I$. Nevertheless, we find that the characteristics of the mean $V$ profile inferred by averaging the subexposures does not differ qualitatively from the $V$ profiles of the individual subexposures. In particular its overall shape and amplitude are not significantly modified. Therefore temporal smearing of the ESPaDOnS and Narval Stokes $V$ and $I$ profiles should not significantly affect either the detectability or the interpretability of any magnetic signatures present in the spectrum of V1449 Aql. 

A related secondary effect involves the depth-dependence of the pulsational dynamics, and therefore a (weak) potential systematic difference in the shape of the line profiles as a function of line strength. To evaluate whether the depth of formation of spectral lines has any influence on the magnetic diagnosis of V1449 Aql, we performed a simple test using Least Squares Deconvolution (LSD; described in detail in Section 3.1) masks filtered for weak (unbroadened central depth $<0.2$ of the continuum) and strong ($>0.2$ of the continuum) spectral lines. The two masks contain similar numbers of spectral lines. We used these masks to extract Stokes $V$ profiles of $\epsilon$~Lupi, the only star of the sample in which a magnetic field is detected. The LSD Stokes $I$ profiles extracted using the two masks are reasonably similar in shape; the asymmetric structure of the line profile (see Fig. 1) is marginally clearer in the ``weak" line, while the wings of the ``strong" line are more extended than those of the ``weak" line. The two LSD profiles differ by about a factor of 1.5 in depth. Longitudinal fields measured from the ``strong" and ``weak" line LSD profiles both correspond to magnetic detections, and agree within the error bars. 

While the two tests described above do not represent a sophisticated evaluation of the temporal smearing and depth-dependence effects on magnetic diagnosis, they illustrate that no significant effect is present at the precision of the ESPaDOnS and Narval observations described in this paper.

In this paper, we present 4 observations of 33 Eri, 4 of 15 CMa, 3 of $\alpha$ Pyx, 2 of $\epsilon$ Lupi, 4 of V1449 Aql and 1 of HY Vel. In the case of HY Vel, poor observing conditions limited the quality of the data and prevented us from coming to any firm conclusions.

The log of CFHT and TBL observations is reported in Table \ref{esp_tab}. For $\epsilon$ Lupi and V1449 Aql, co-added spectra were created from observations taken on the same night in order to increase the SNR. In the case of V1449 Aql radial velocity differences (described above) between observations were significant enough that spectra were brought to a common velocity before being combined. 

\begin{table*}

 \begin{minipage}{180mm}
\begin{center}
  \caption{Summary of results from ESPaDOnS and Narval data. Note that subexposure times correspond to 1/4 of the total exposure time.}
  \begin{tabular}{@{}lcccccrcrrcrc@{}}
  \hline
   Star & HD & Inst. & Exp. & Number & HJD & Peak & Phase  & Range & $\langle  B_z\rangle $ & Det. Prob. & $\langle N_z\rangle$ & Det. Prob.\\
   & Number & & Time (s) & Spectra &- 2400000 & S/N & & (km~s$^{-1}$) & (G) & ($V$) & (G) &  ($N$)\\
 \hline
$\epsilon$ Lupi & 136504 & E & 80 & 4 & 55634.1488 & 1088 & -- & -79 -- 85 & \textbf{-90$\pm$17} & 1.000 & -2$\pm$17 & 0.001 \\
& & E & 80 & 4 & 55727.8071 & 1241 & -- & -79 -- 85 & \textbf{-76$\pm$16} & 1.000 & -20$\pm$16 & 0.011\\
$\alpha$ Pyx & 74575 & E & 600 & 1 & 55549.0736 & 1202 & 0.643 &  -40 -- 70 & -5$\pm$6 & 0.016 &  -4$\pm$6 & 0.053  \\
& &	E & 600 & 1 & 55555.0931 & 1270 & 0.525 & -40 -- 70 & -8$\pm$6 & 0.246 & -1$\pm$6 & 0.236\\
& &	E & 600 & 1 & 55561.0935 & 1320 & 0.402 & -40 -- 70 & 11$\pm$5 & 0.693 & 4$\pm$5 & 0.005\\
33 Eri & 24587 & N & 1200 & 1 & 54373.9793 & 1105 & 0.316 & -45 -- 80 & -46$\pm$48 & 0.341 & -4$\pm$48 & 0.004\\ 
& & E & 600 & 1 & 55547.7521 & 809 & 0.074 &  -55 -- 50 & 35$\pm$25 & 0.001 &  -9$\pm$25 & 0.055  \\
& & E & 600 & 1 & 55549.9444 & 917 & 0.787 & -55 -- 50 & -19$\pm$22 & 0.001 & 9$\pm$22 & 0.018\\
& &	E & 600 & 1 & 55554.8367 & 745 & 0.611 & -55 -- 50 & 4$\pm$34 & 0.290 & -15$\pm$34 & 0.015\\
15 CMa & 50707 & E & 600 & 1 & 55250.9147 & 844 & 0.676 &  -50 -- 100 & -13$\pm$13 & 0.353 &  4$\pm$13 & 0.045 \\
& & E & 600 & 1 & 55549.0566 & 869 & 0.261 & -50 -- 100 & -38$\pm$19 & 0.000 & -15$\pm$19 & 0.007\\
& &	E & 600 & 1 & 55555.0197 & 785 & 0.733 & -50 -- 100 & -4$\pm$14 & 0.003 & 13$\pm$14 & 0.189\\
& & E & 600 & 1 & 55557.9927 & 686 & 0.968 & -50 -- 100 &  -3$\pm$16 & 0.022 & 11$\pm$16 & 0.002\\
HY Vel & 74560 & E & 240 & 1 & 55611.9666 & 460 & -- & -40 -- 60 & 22$\pm$217 & 0.750 & 4$\pm$217 & 0.303\\
V1449 Aql & 180642 & E & 2360 & 2 & 55721.9664 & 326 & 0.998 & -90 -- 90 & 28$\pm$53 & 0.000 & -91$\pm$53 & 0.307 \\
& & E & 960 & 3 & 55749.8108 & 441 & 0.002 & -120 -- 60 & 26$\pm$53 & 0.009 & -10$\pm$53 & 0.103 \\
& & E & 960 & 3 & 55756.0395 & 826 & 0.450 & -100 -- 80 & 2$\pm$30 & 0.001 & -11$\pm$30 & 0.008 \\
& & N & 960 & 6 & 55783.5336 & 847 & 0.429 & -100 -- 80 & 30$\pm$50 & 0.000 & -18$\pm$49 & 0.094 \\
\hline
\label{esp_tab}
\end{tabular}
\end{center}
\end{minipage}

\end{table*}

\subsection{FORS1/2 and SOFIN}

We have re-analyzed the FORS1/2 observations\footnote{After decommissioning in March 2009, the polarimetric optics of FORS1 were moved to FORS2.} obtained by H2011a, both in published and re-reduced forms. H2011a presented 62 Stokes $V$ observations of 5 of our program stars, obtained between November 2003 and March 2010 using grism 600B, delivering a spectral resolution R$\sim$2000. The observations sample this time period nonuniformly, and the number of observations varies significantly from star to star (7 observations each for $\alpha$~Pyx and $\epsilon$~Lup, 15 observations each for 15 CMa and HY Vel, and 18 observations for 33 Eri). 

We also analyzed the 2 FORS1/2 and 13 SOFIN measurements of V1449 Aql reported by H2011b. The SOFIN spectra were obtained from August--September 2009 and in July 2010, and were made at intermediate spectral resolution (R$\sim$30,000).

\section{Measurements of Longitudinal Magnetic Fields}

\subsection{Measurements from ESPaDOnS and Narval spectra}

The ESPaDOnS/Narval spectra were analyzed using the Least Squares-Deconvolution (LSD) multiline analysis method developed by Donati et al. (1997). LSD combines the information from essentially all metallic and He lines in the spectrum by means of the assumption that the spectrum can be reproduced by the convolution of a ``mean" line profile (the ``LSD profile") with an underlying spectrum of unbroadened atomic lines of given line depth, Land\'e factor, and wavelength (the ``line mask", as described by Wade et al., 2000). This process allows the computation of averaged Stokes \textit{I}, \textit{V} and \textit{N} profiles with much higher signal-to-noise ratios than those of the individual lines, dramatically improving the detectability of Zeeman signatures due to stellar magnetic fields. The application of LSD to spectra of pulsating B stars is discussed in detail by Silvester et al. (2009).

The line masks used in this analysis were modified from generic masks obtained using the Vienna Atomic Line Database (VALD; Piskunov et al., 1995; Ryabchikova et al., 1997; Kupka et al., 1999; Kupka et al., 2000) {\sc extract stellar} requests. The VALD line lists are characterized by the stellar effective temperatures and gravities of the individual stars (as reported by H2011a), solar abundances, and a minimum line depth of 10\% of the continuum. The masks were customized to the spectrum of each star by interactive comparison, removing weak lines and those blended with Balmer or telluric lines. This `cleaning' of the mask ensures that lines included in the LSD procedure have shapes that are as similar and uniform as possible. 

\begin{figure*}
\includegraphics[width = 6cm]{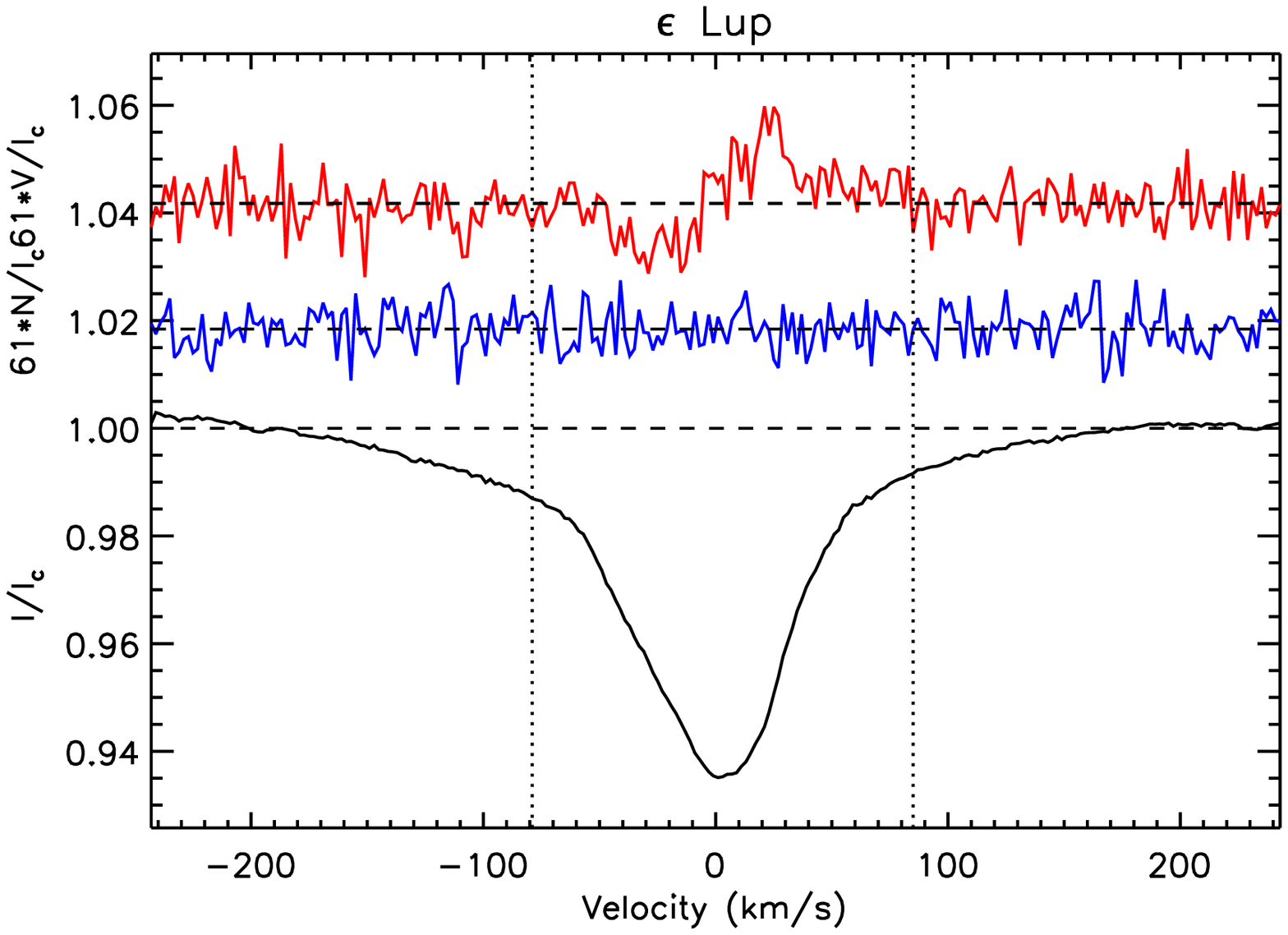}
\includegraphics[width = 6cm]{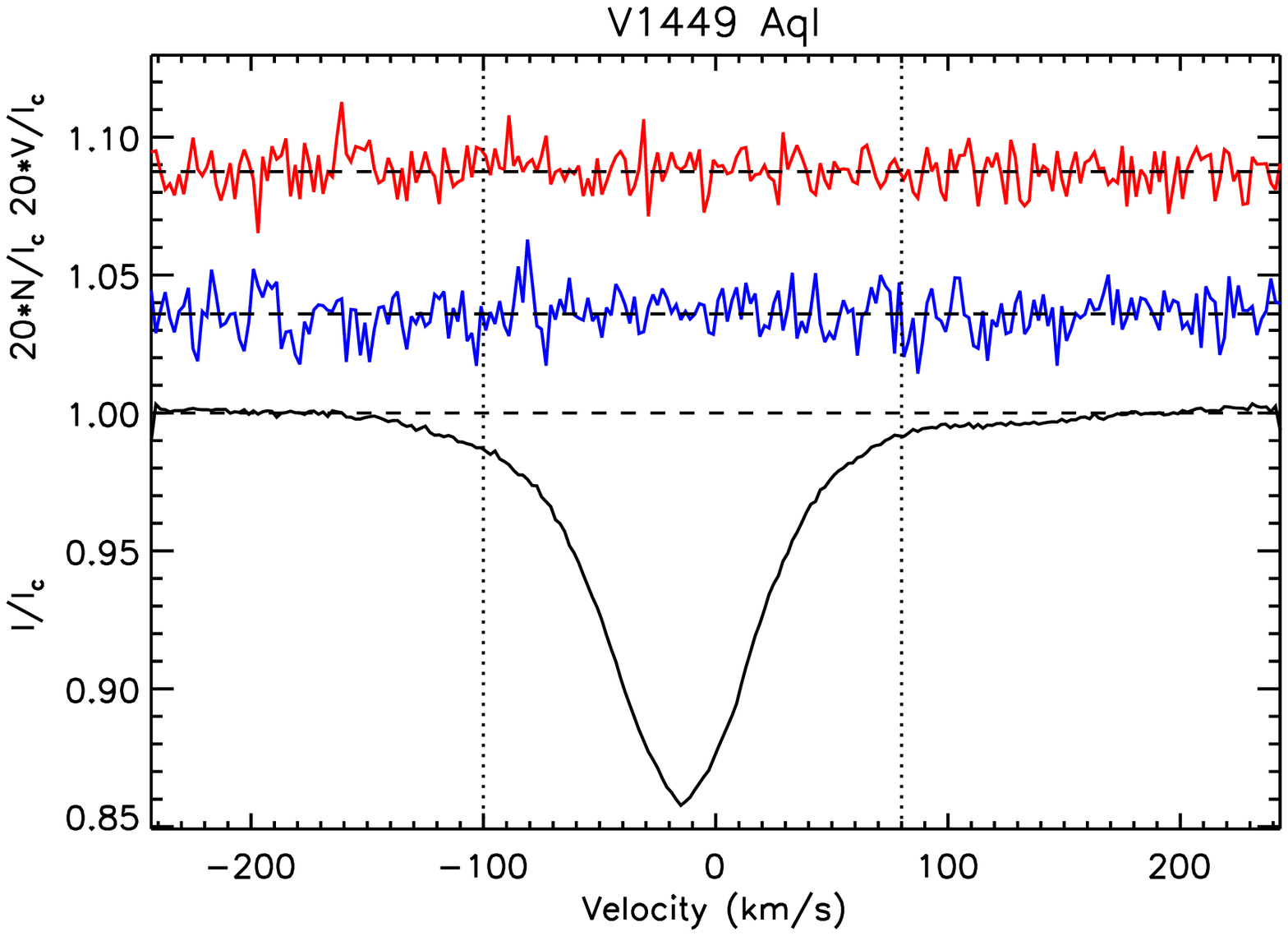}
\includegraphics[width = 6cm]{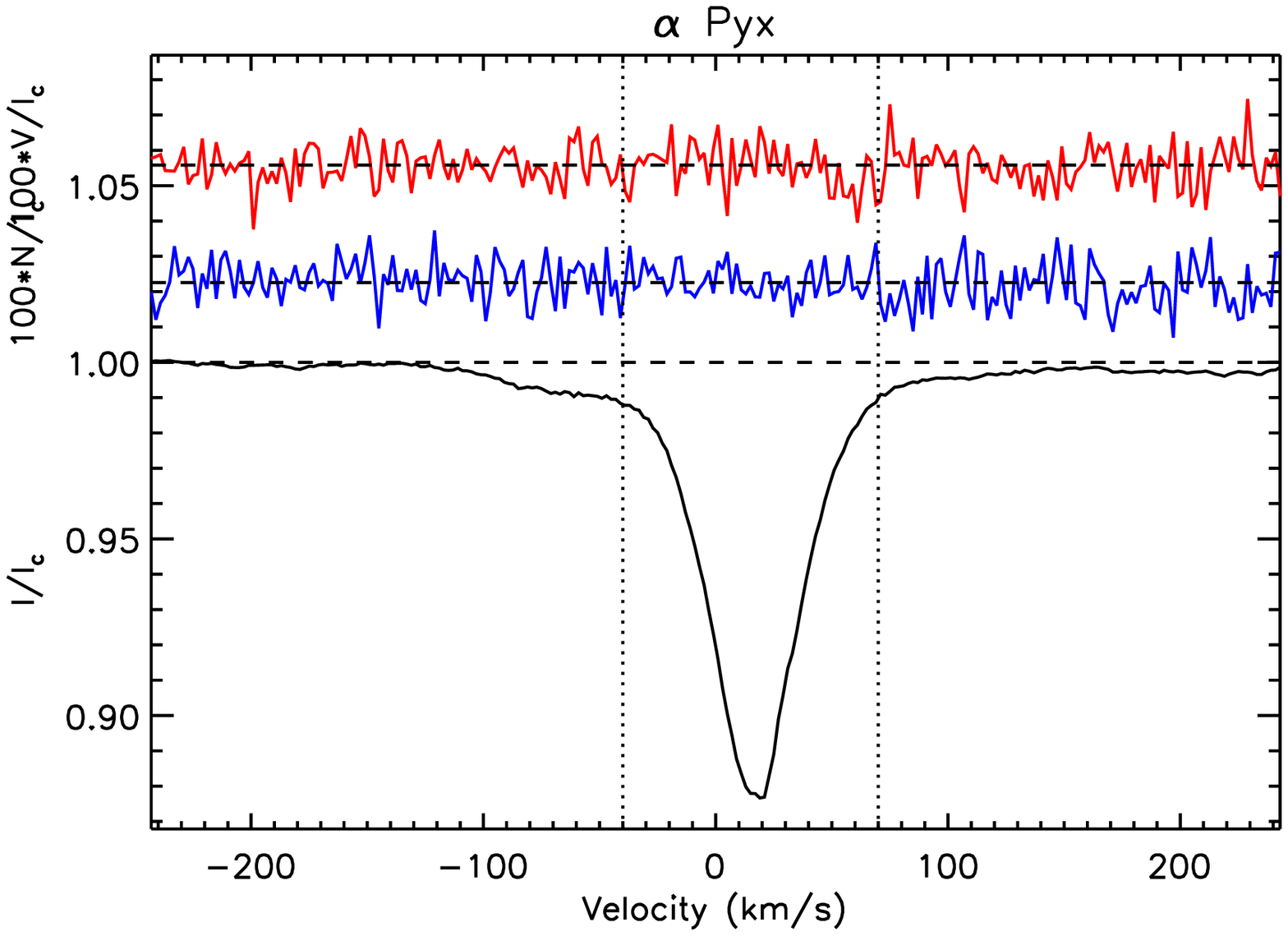}
\includegraphics[width = 6cm]{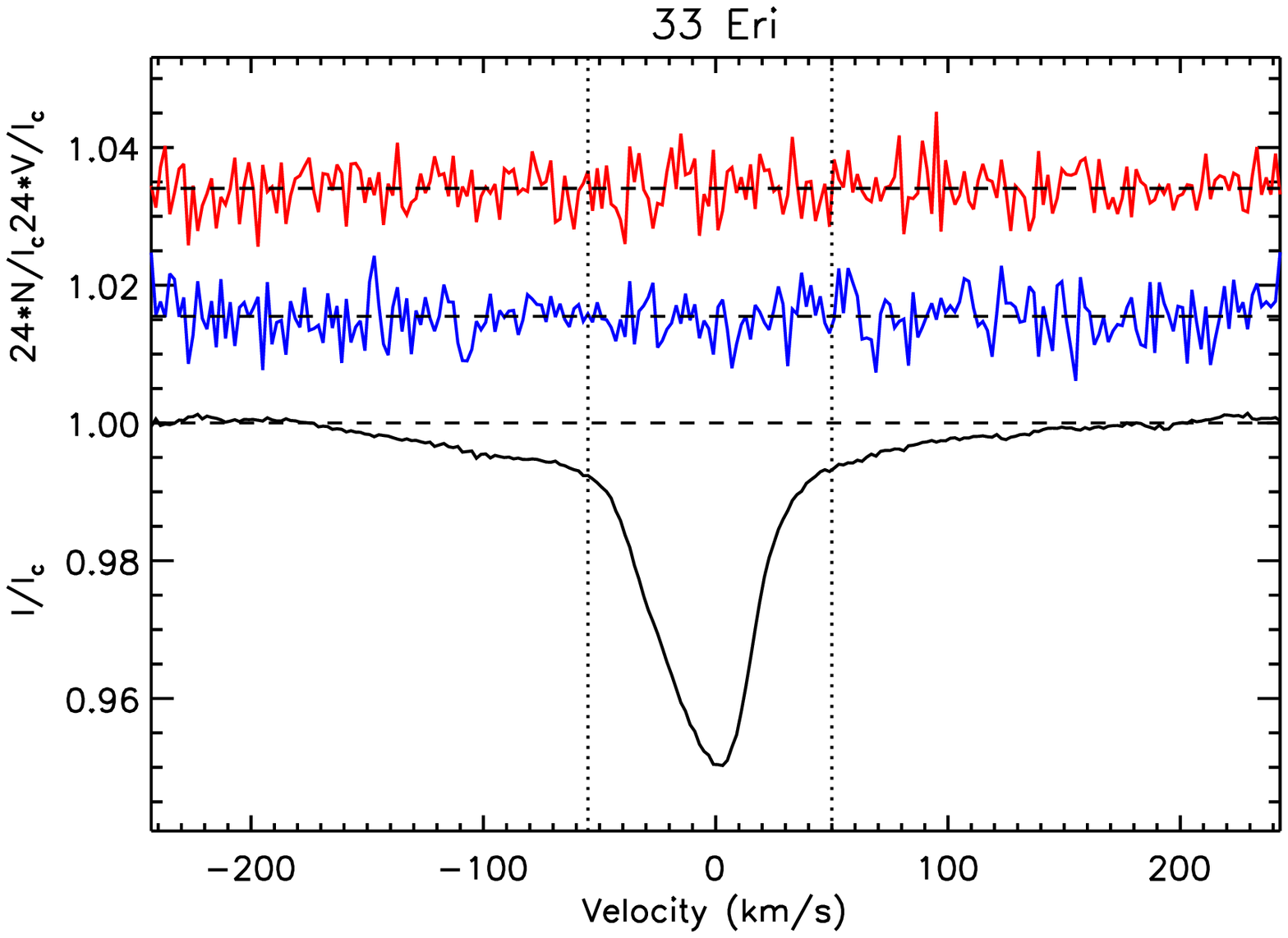}
\includegraphics[width = 6cm]{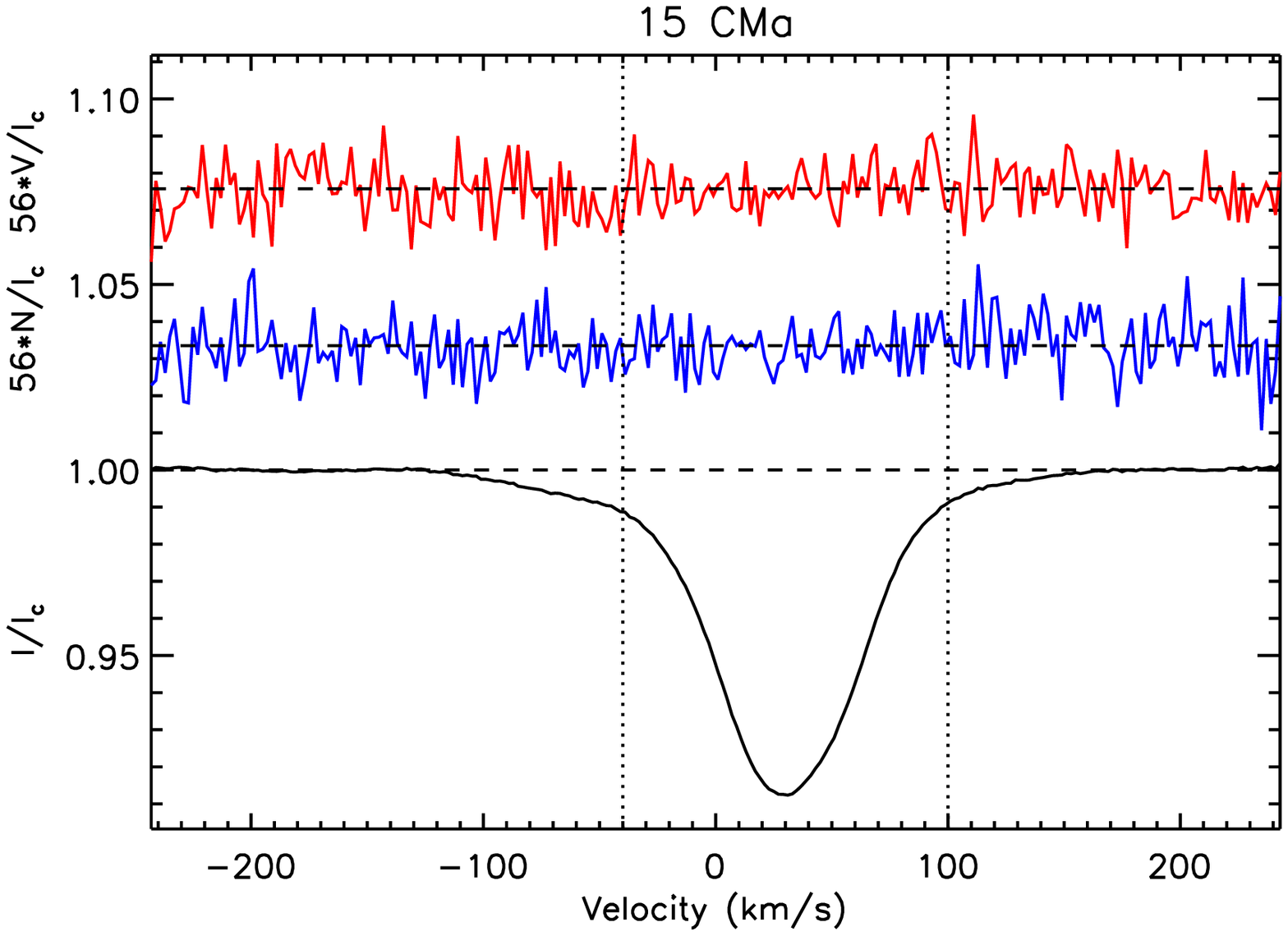}
\includegraphics[width = 6cm]{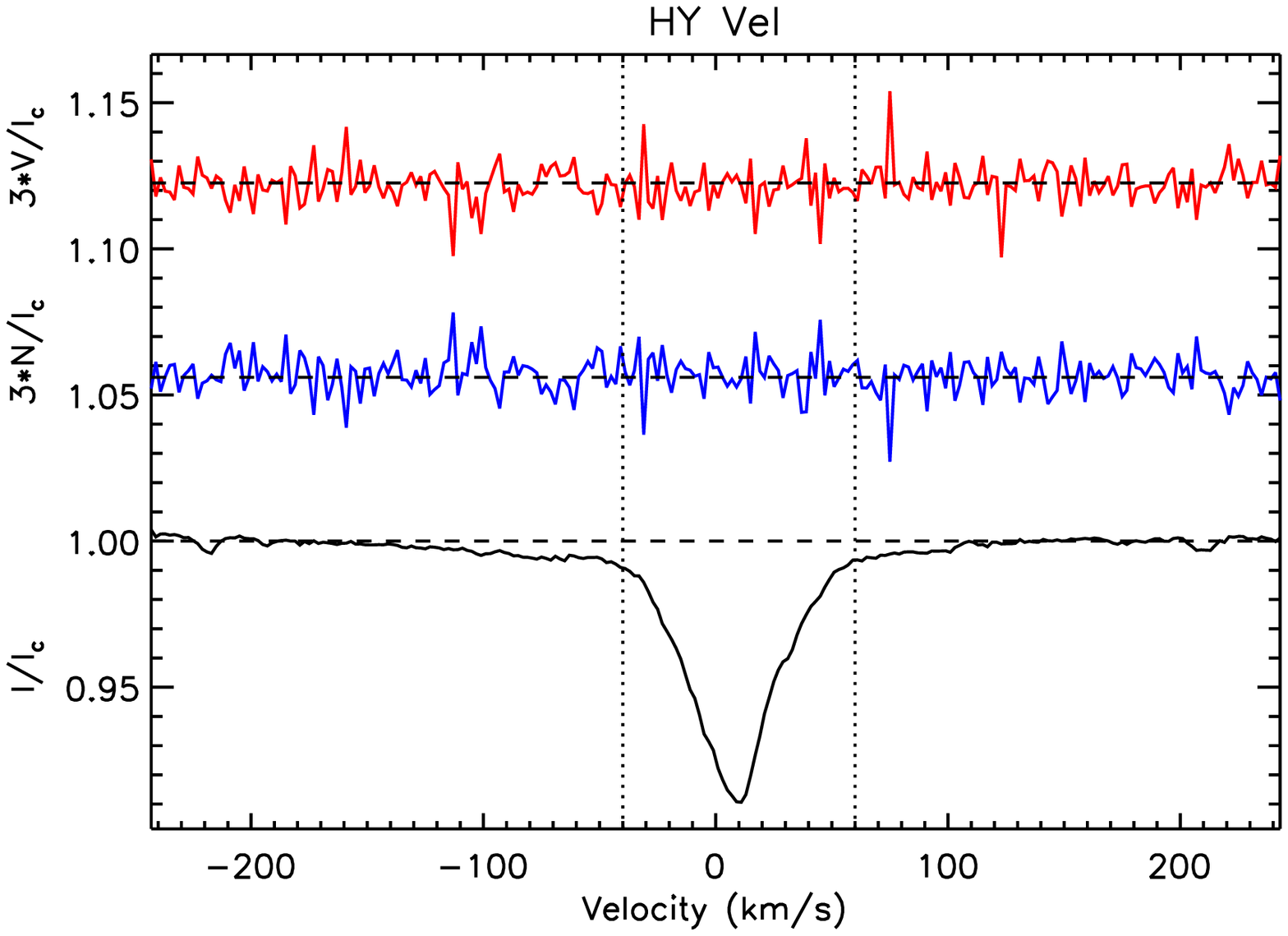}
\caption{Examples of LSD profiles of stars observed with ESPaDOnS and Narval. Stokes \textit{I} in black (bottom), diagnostic \textit{N} in blue (middle), \textit{V} in red (top). Both $V$ and $N$ have been scaled and shifted for display purposes. Vertical lines denote the integration range adopted for longitudinal field measurement. \textit{Top, left--right}: $\epsilon$ Lupi, V1449 Aql,  $\alpha$ Pyx, \textit{Bottom, left--right}: 33 Eri, 15 CMa, HY Vel. Note the similarity between $N$ and $V$ in 5 of the 6 stars. Note also that the scaling of $V$ and $N$ varies substantially between stars. For each star, the LSD profile with the highest SNR is shown.}
\label{LSD} 
\end{figure*}

\begin{figure*}
\begin{tabular}{cc}
\includegraphics[width = 9.cm]{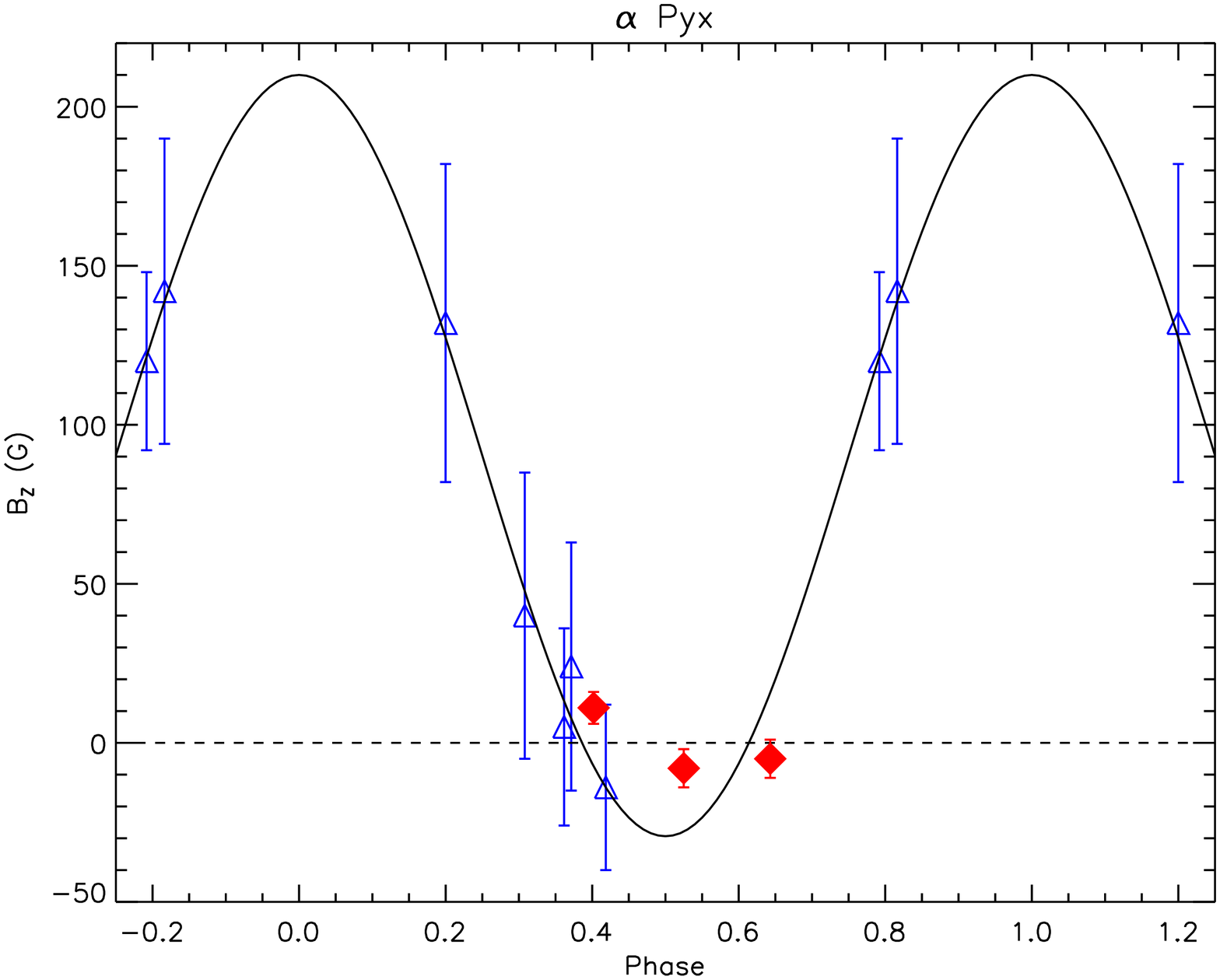} & 
\includegraphics[width = 9.cm]{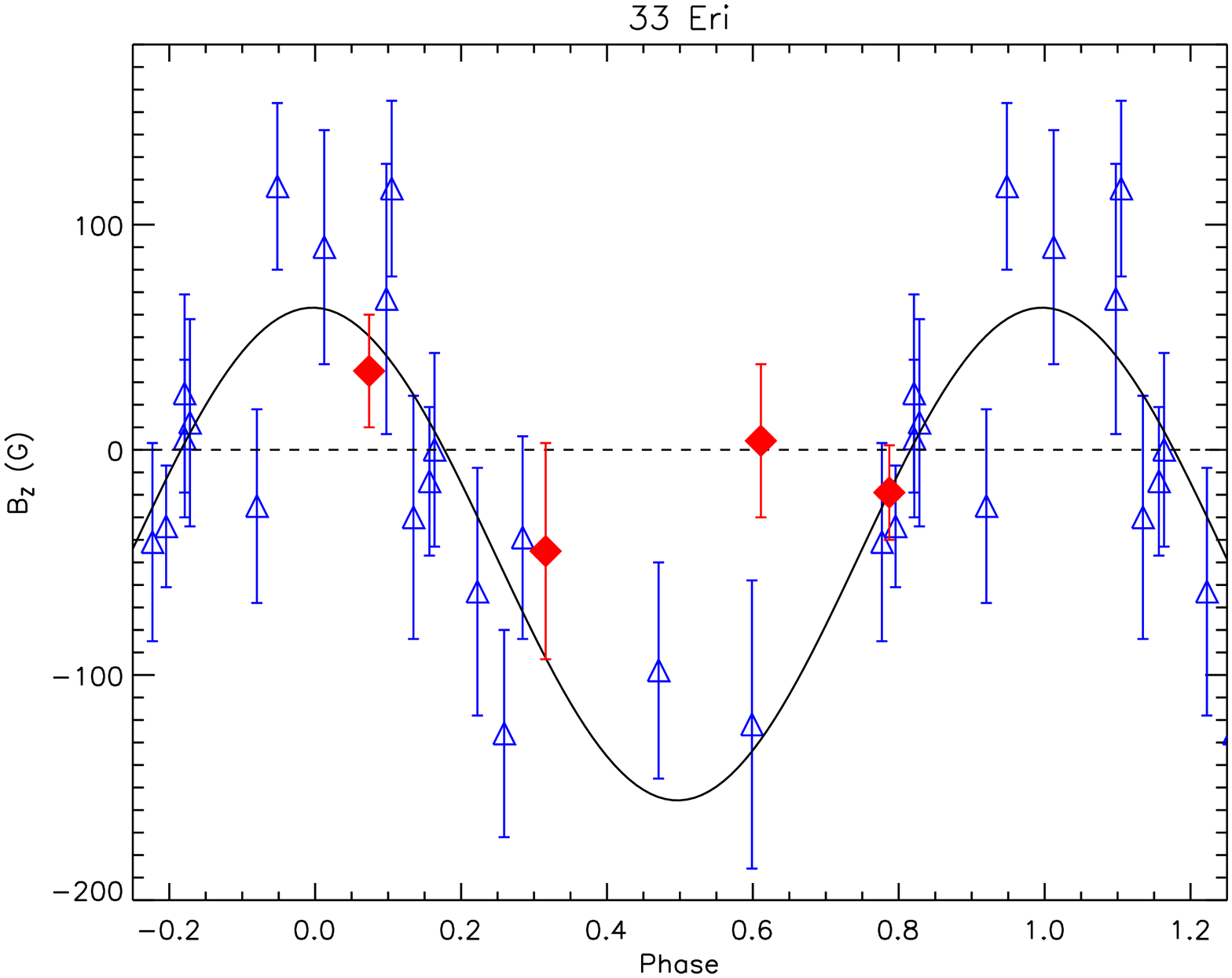} \\
\includegraphics[width = 9.cm]{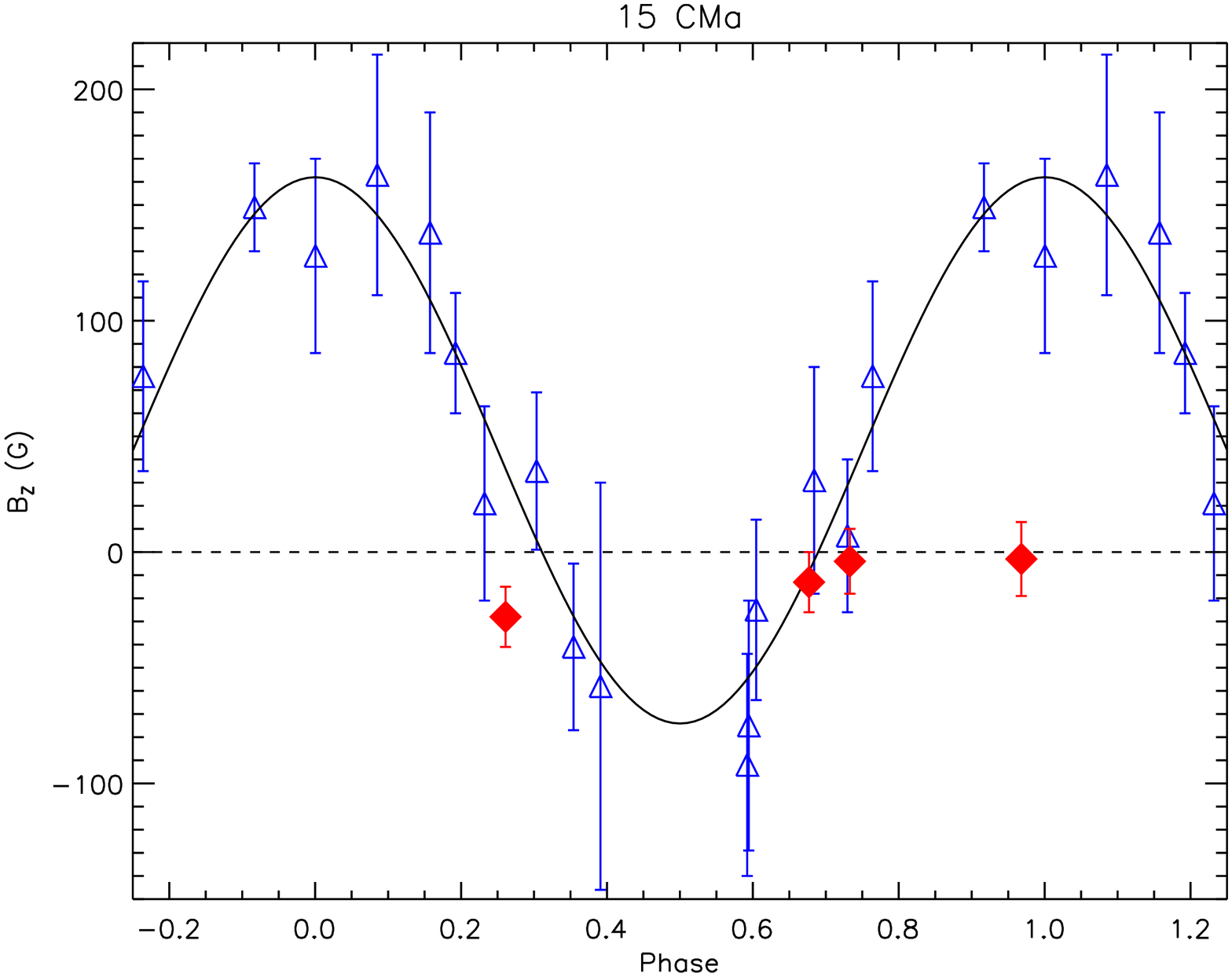} &
\includegraphics[width = 9.cm]{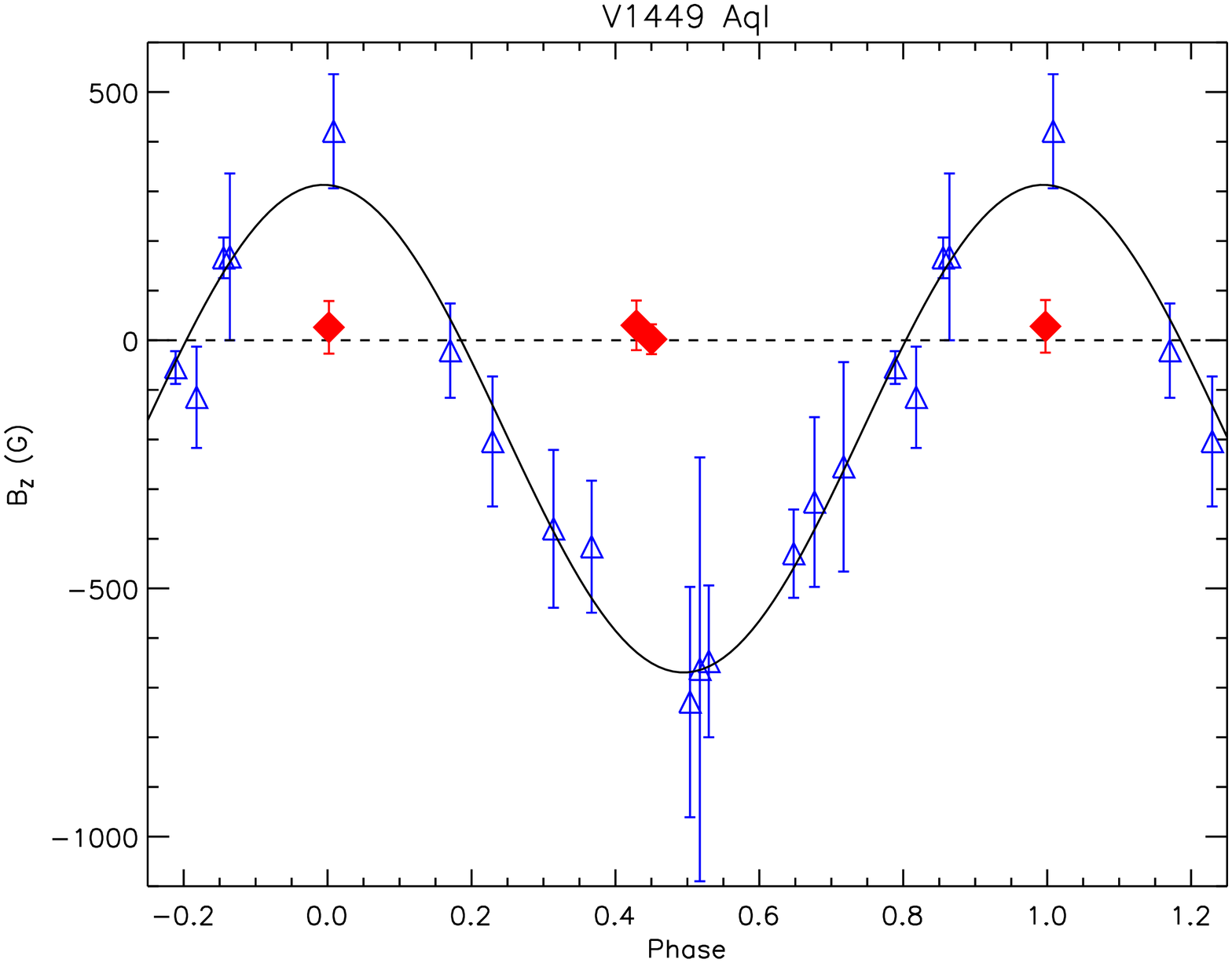} \\
\end{tabular}
\caption{Longitudinal magnetic field measurements obtained with FORS1/2 (open blue triangles, from H2011a and H2011b) and ESPaDOnS/Narval (filled red diamonds). The data have been phased according to the ephemerides proposed by H2011a and H2011b. All ESPaDOnS/Narval measurements are consistent with the absence of magnetic field: the dotted line shows $\langle  B_z\rangle =0$, while the solid line shows the field curve from the H2011a and H2011b models. \textit{Top--Bottom}: $\alpha$ Pyx, 33 Eri, 15 CMa, V1449 Aql}
\label{fors_esp_com} 
\end{figure*}

The lines remaining in the mask following the cleaning procedure -- between 230 and 320 lines depending on the physical and spectral properties of the star -- were then adjusted to better match the observed line depths in the associated observed spectrum. `Tweaking' the lines in this fashion improves the fit between the actual spectrum and the LSD spectrum model (that is, the convolution of the LSD profile and the line mask). 

Two methods were used to evaluate the presence of a photospheric magnetic field: a statistical test performed on the Stokes $V$ profile, and direct inference based on the significant detection of a longitudinal magnetic field. The former method, described by Donati, Semel, \& Rees (1992) and Donati et al. (1997), employs the reduced $\chi^2$ of the signal in Stokes \textit{V} within the bounds of the line profile. It reports the detection of a magnetic signature as `definite' if the formal detection probability is greater than 99.999\%, while the detection probabilities outside the Stokes \textit{V} line profile, and inside the diagnostic null (\textit{N}) line profile, are both negligible; a 'marginal' detection corresponds to a detection probability between 99.9\% and 99.999\%. 

The second, direct inference method involves computation of the mean longitudinal magnetic field from the first-order moment of the Stokes \textit{V} profile within the line according to the expression: 

\begin{equation}
\langle B_z  \rangle  = -2.14\times 10^{11}\frac{\int \! vV(v)\mathrm{d}v}{\lambda g_{\rm{eff}}c\int \left[I_{\rm{c}}-I(v)\right] \mathrm{d}v}
\end{equation}

\noindent where $\langle  B_z\rangle$ is the mean longitudinal magnetic field in G, \textit{v} is the velocity in km~s$^{-1}$ within the profile measured relative to the centre of gravity (Mathys et al., 1989; Donati et al. 1997; Wade et al., 2000), and $\lambda$ and $g_{\rm eff}$ are the reference values of the wavelength in nm and Land\'e factor used in computing the LSD profiles. This equation is also applied to the null profile, yielding a comparable measurement $\langle N_z\rangle$. The uncertainties associated with $\langle  B_z\rangle$ and $\langle  N_z\rangle$ were determined by propagating the formal (photon statistical) uncertainties of each pixel through Eq. (1). This is described in more detail by Silvester et al. (2009). 

The integration ranges employed in the evaluation of Eq. (1) associated with each LSD profile were selected individually through visual inspection so as to include the entire span of the Stokes \textit{I} profile. These ranges are indicated in Fig. \ref{LSD}, and reported for each observation in Table \ref{esp_tab}. Integration ranges can differ significantly for different observations of the same star, especially when pulsations are a factor as they are here, and the ability to tailor the integration range to the shape of each line profile and thus optimize the magnetic diagnosis is an important strength of high-resolution spectropolarimetry. The pulsational character of the stars is clearly visible in the often rather asymmetric LSD Stokes \textit{I} profiles. We tested the sensitivity of the inferred longitudinal field and its error bar to the integration range by adjusting the integration limits by $\pm 10\%$. This has the effect of decreasing (for smaller range) or increasing (for larger range) the error bar by 15-20\%, but otherwise has no impact on the results. These results are qualitatively consistent with experiments carried out by Neiner et al. (2012) (see their Fig. 3).

Most stars show relatively broad wings, attributable to the inclusion of He lines in the line masks. Removing the He lines from the mask reduces this effect at the expense of sensitivity, yielding an increase in the error bars of $\langle  B_z\rangle $ on order of 50\%. Experiments using $\epsilon$~Lupi (the only star of the sample in which magnetic field is detected in the ESPaDOnS/Narval spectra) clearly show that masks including and excluding He lines yield congruent results in the presence of a magnetic field (apart from differences in error bars). The line masks from which the LSD profiles shown in Fig. \ref{LSD} were constructed include He lines.

\subsection{Published measurements}

The mean longitudinal magnetic field $\langle  B_z\rangle $ of FORS1/2 data was derived using the linear regression method developed by Bagnulo et al. (2002). The measurements were obtained in two ways: using only hydrogen Balmer lines and using the full spectrum (including H Balmer lines along with metal and He lines). No diagnostic null measurements were reported. We assume that uncertainties were derived from the formal uncertainties obtained from the regression. These measurements, which we re-analyze below, are reported in Tables 2 and 3 of H2011a. 

$\langle  B_z\rangle $ was measured from SOFIN using the moment technique of Mathys (1994). Again no diagnostic nulls are reported. The measurements are provided in Table 1 of H2011b. 

\subsection{Measurements from re-reduced spectra}

In order to test the robustness of the field measurements reported by H2011a, the archival data from which they were determined were obtained and re-reduced using a new FORS reduction pipeline, the details of which are described by Bagnulo et al. (2012). We have measured the longitudinal field using the method of Bagnulo et al. (2002) from H lines (H$\beta$ to the Balmer jump), from the full spectrum (from 3800~\AA--4950~\AA), as well as from ``metal lines" (i.e. the full spectrum excluding H lines). The measurements were performed using a larger aperture for spectrum extraction, and omitting the region 3900--4000 \AA, which has been found to be afflicted by internal reflections. For all three sets of measurements obtained from the Stokes $V$ spectra, longitudinal fields from the corresponding diagnostic null spectra were also measured. Full spectrum and H line measurements are compared to the original published measurements in Fig. \ref{rered_compare}.

At the time of this writing the SOFIN data remains proprietary and so a re-reduction cannot be undertaken.


\section{Results}


The results of the LSD analysis are summarized in Table \ref{esp_tab}, and illustrated in Figs. \ref{LSD} and \ref{fors_esp_com}. According to our detection criteria, just one of the 6 stars -- $\epsilon$ Lupi -- shows evidence for the presence of a magnetic field in its photosphere. $\epsilon$~Lup exhibits a significant Stokes $V$ LSD profile, corresponding to longitudinal magnetic fields detected at 5.3$\sigma$ and 4.8$\sigma$. The remaining five stars exhibit no significant longitudinal magnetic field nor Stokes $V$ signal.

To evaluate the robustness of the magnetic field diagnosis from Stokes $V$, we have also computed both the detection probability and longitudinal field using the $N$ LSD profiles. In no case is any significant polarised signal or longitudinal field detected in the those data.

\textit{The remarkable result of the ESPaDOnS and Narval observations is the absence of any evidence of magnetic field for four stars -- $\alpha$ Pyx, 15 CMa, 33 Eri and V1449 Aql -- for which magnetic field geometry models were presented by H2011a and H2011b.}

\begin{figure}[ht]
\centering
\includegraphics[width=4.25cm]{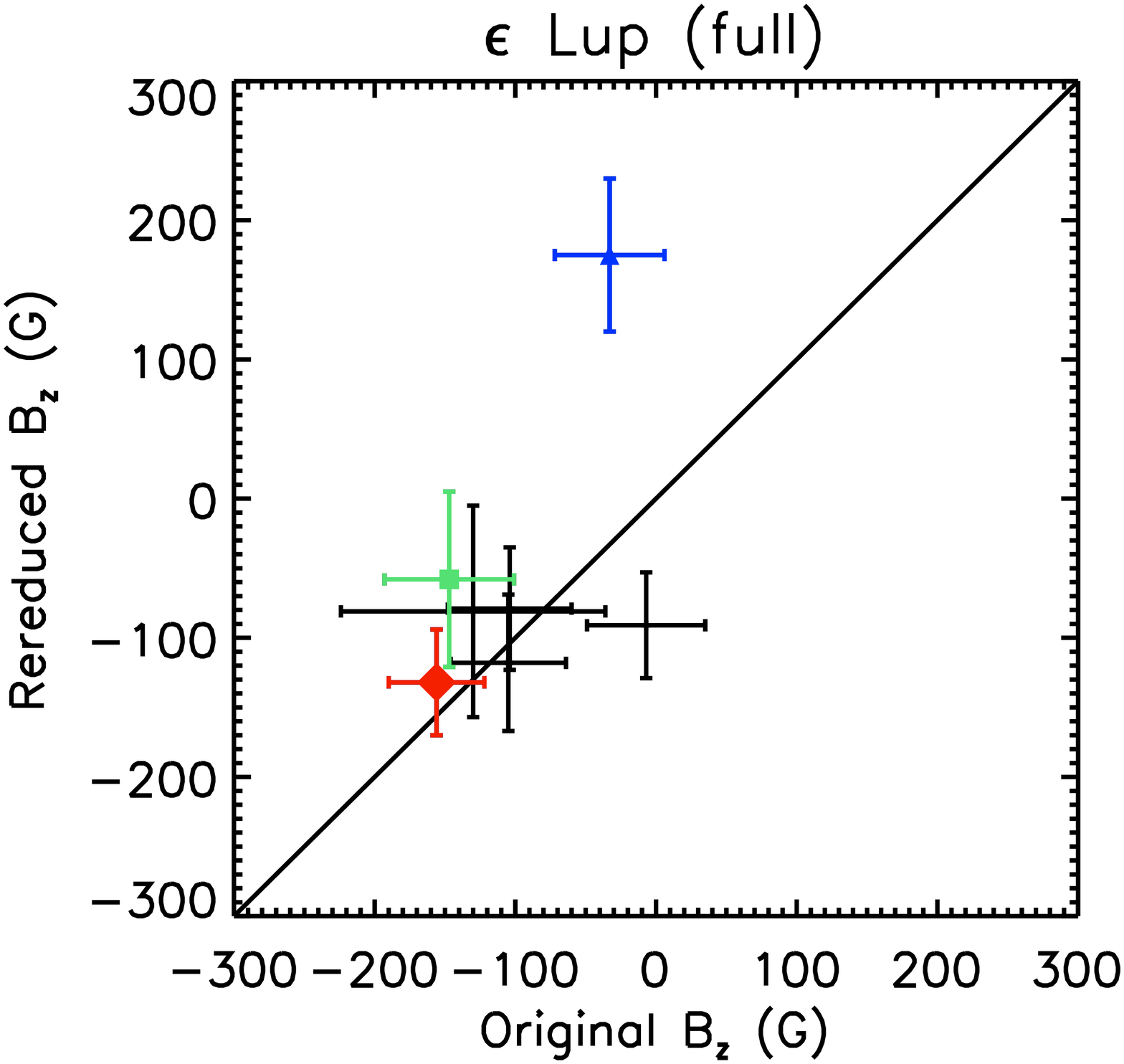}
\includegraphics[width=4.25cm]{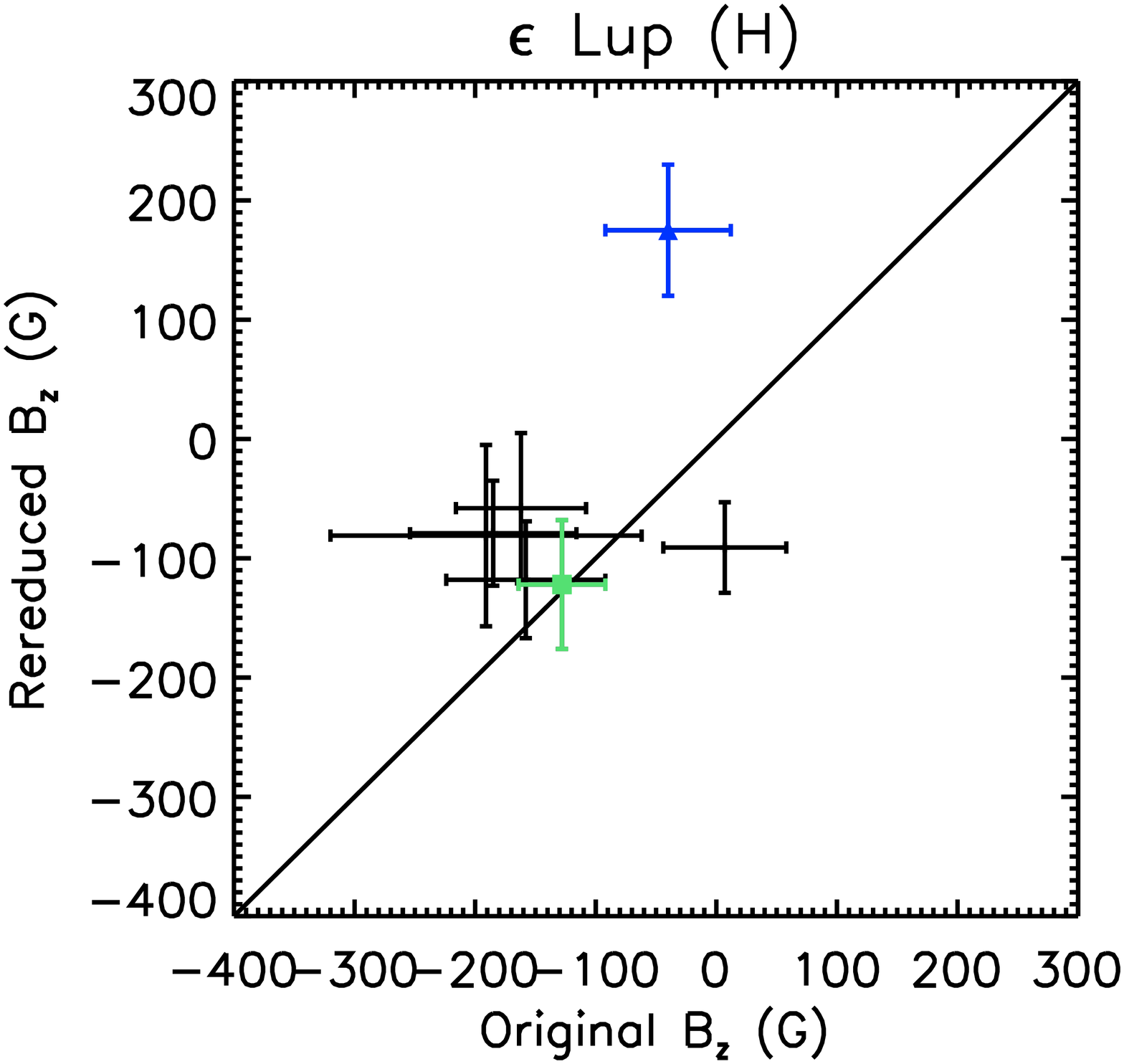}
\includegraphics[width=4.25cm]{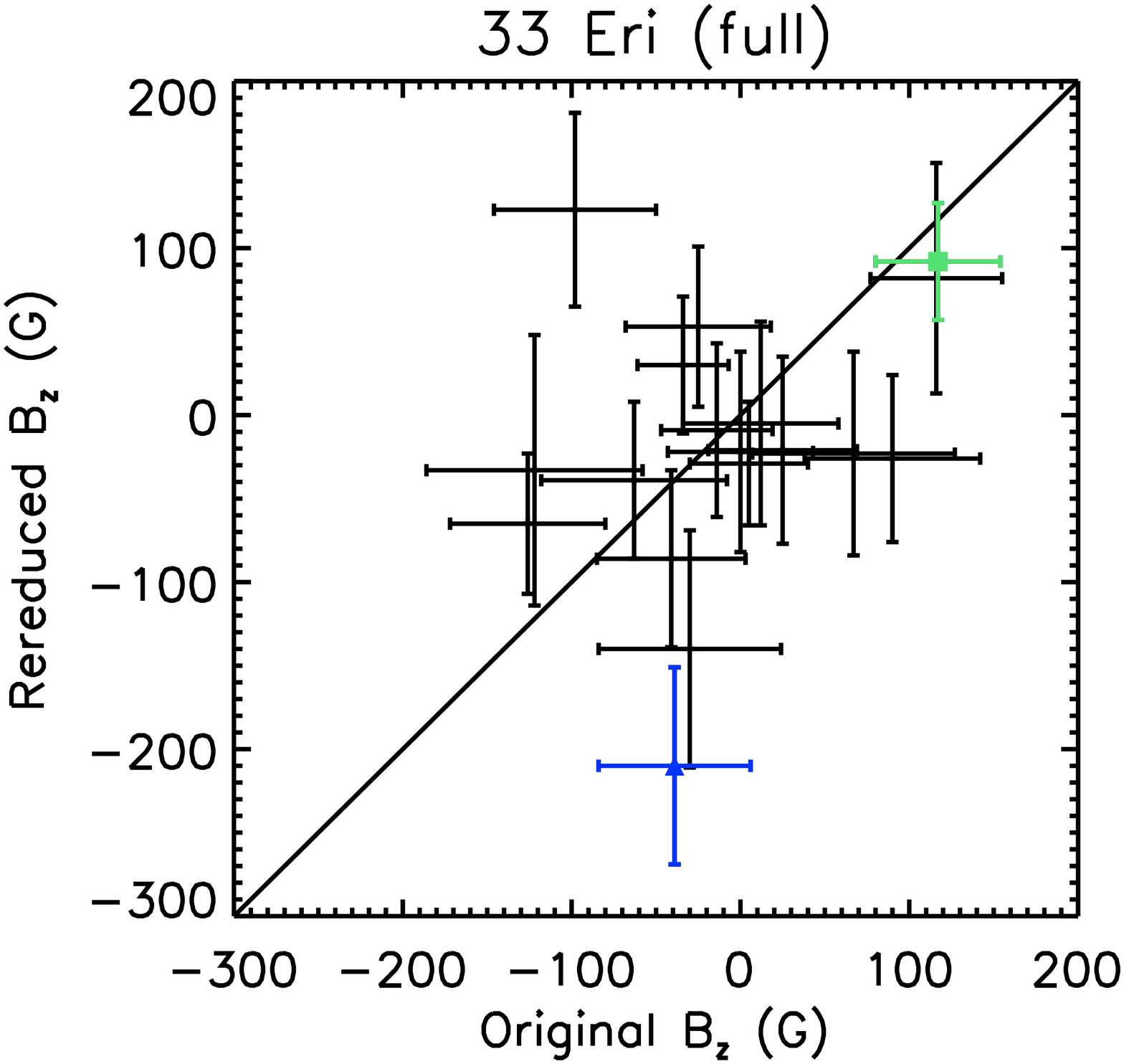}
\includegraphics[width=4.25cm]{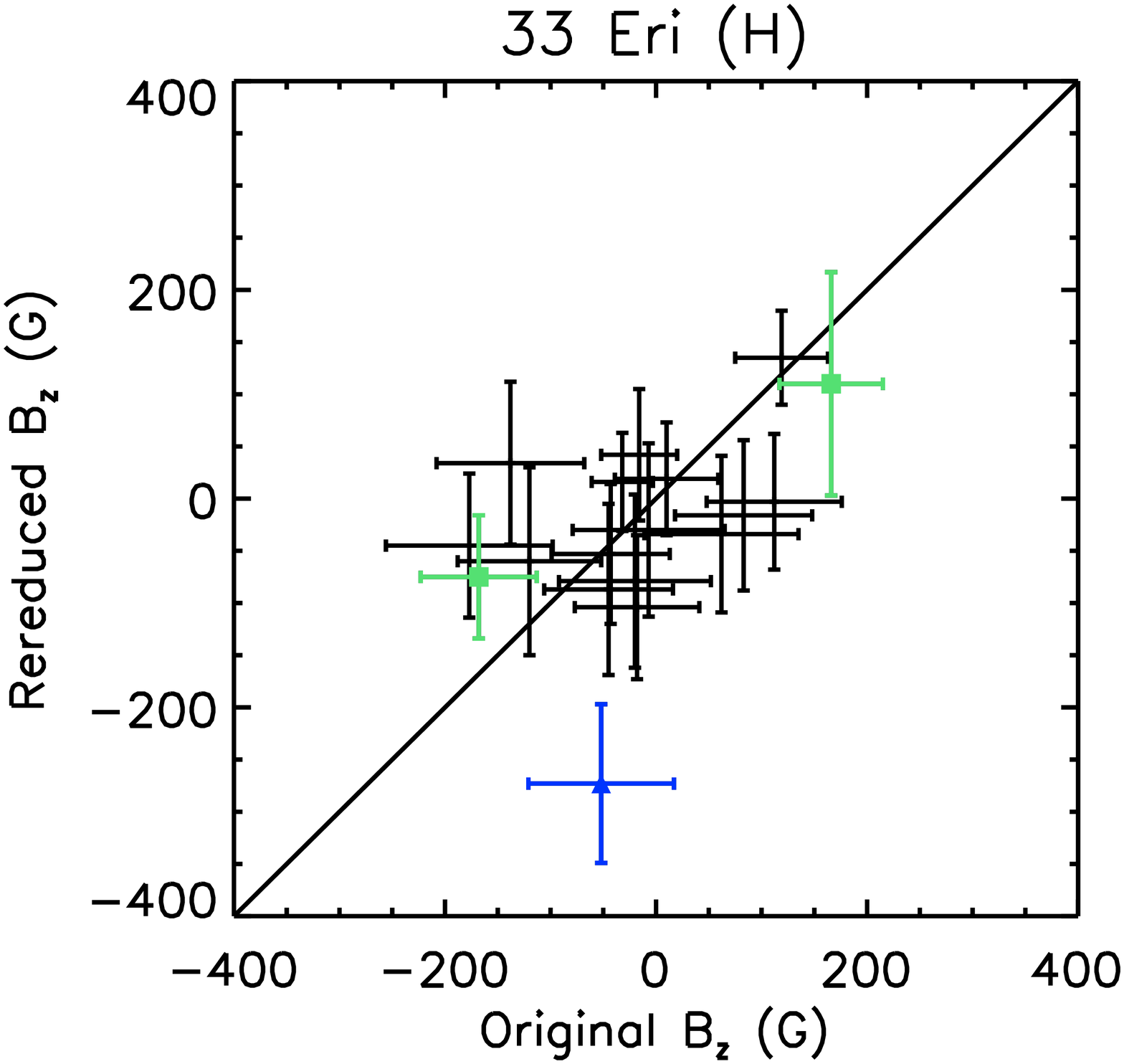}
\includegraphics[width=4.25cm]{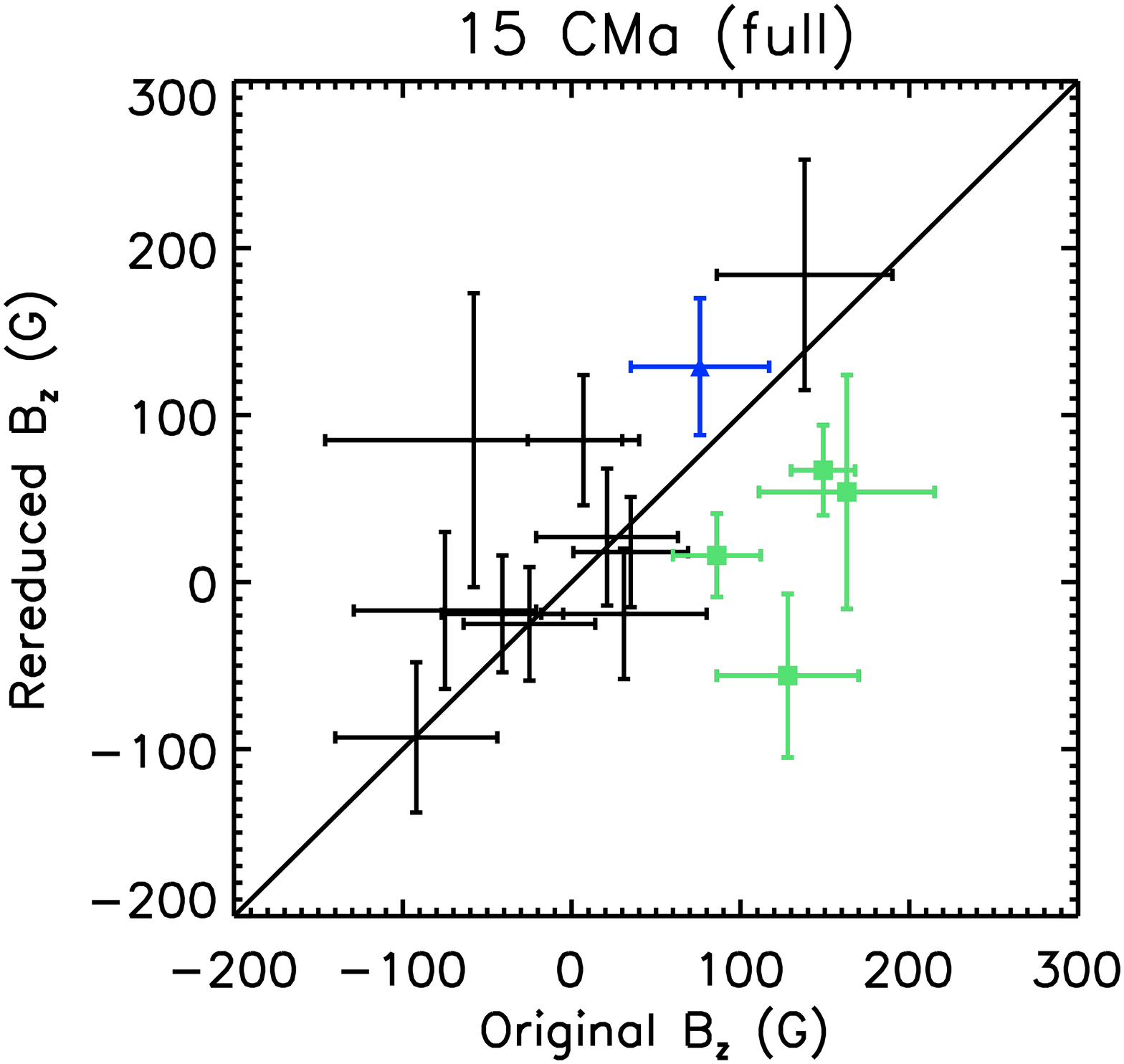}
\includegraphics[width=4.25cm]{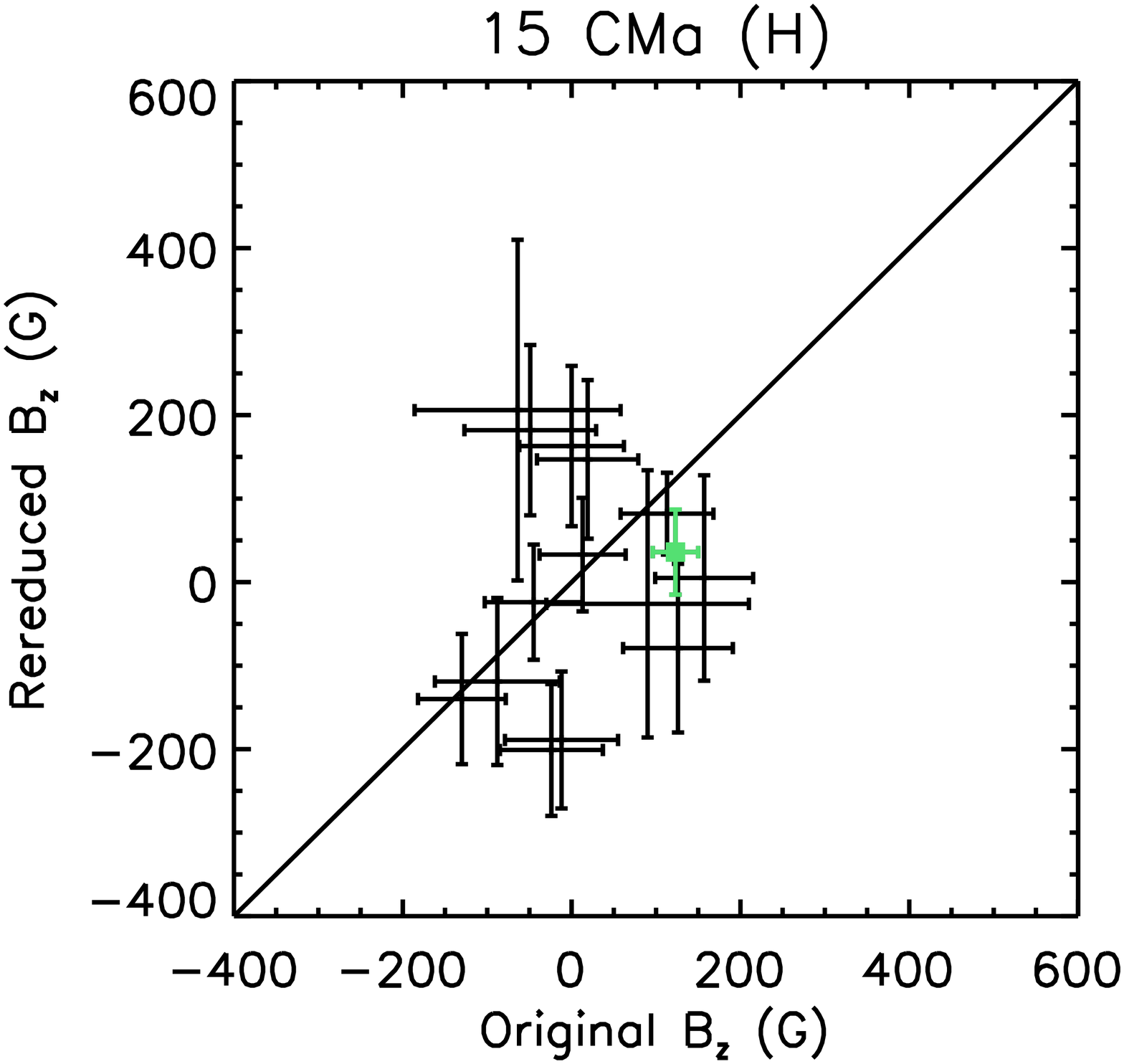}
\includegraphics[width=4.25cm]{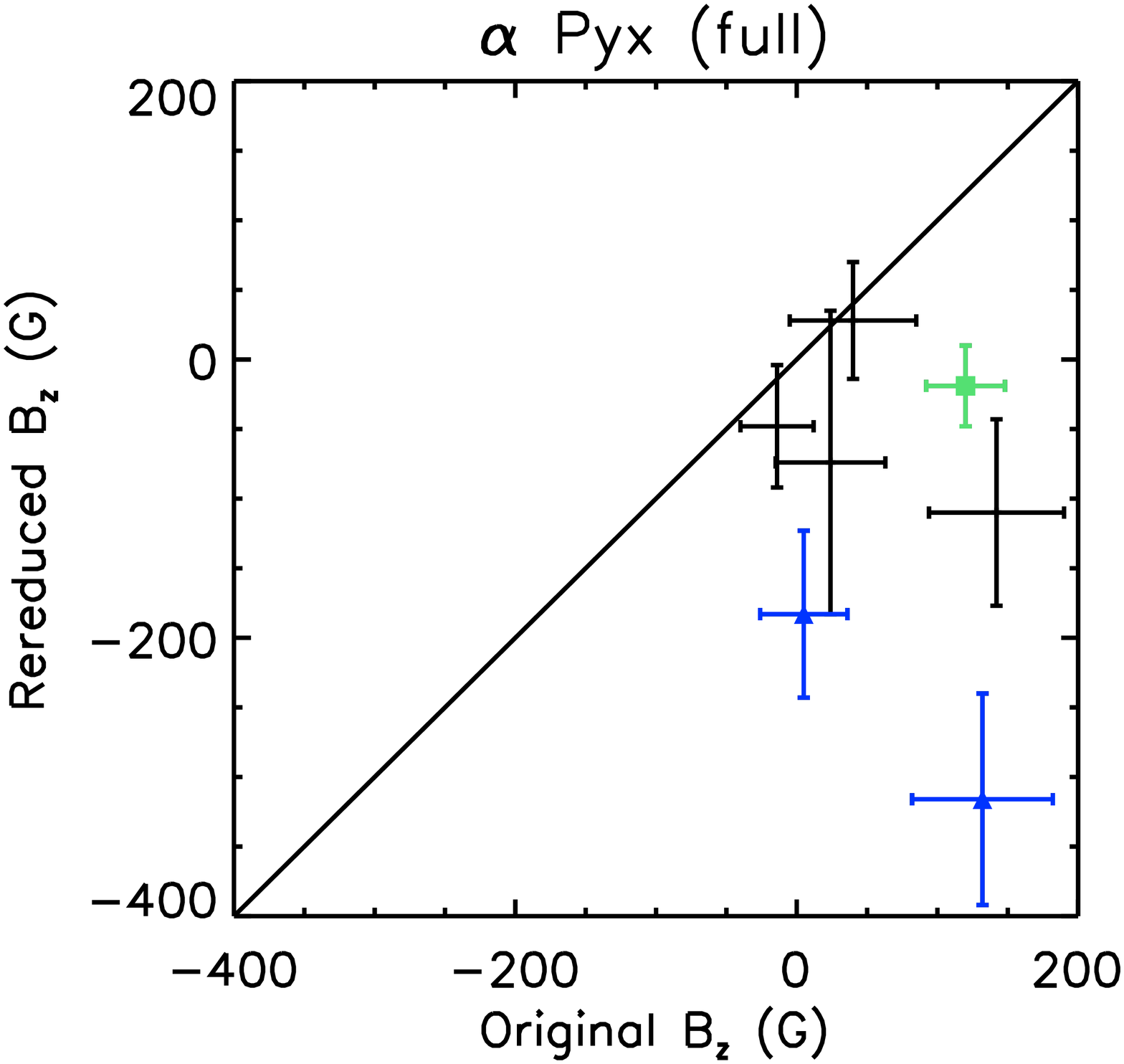}
\includegraphics[width=4.25cm]{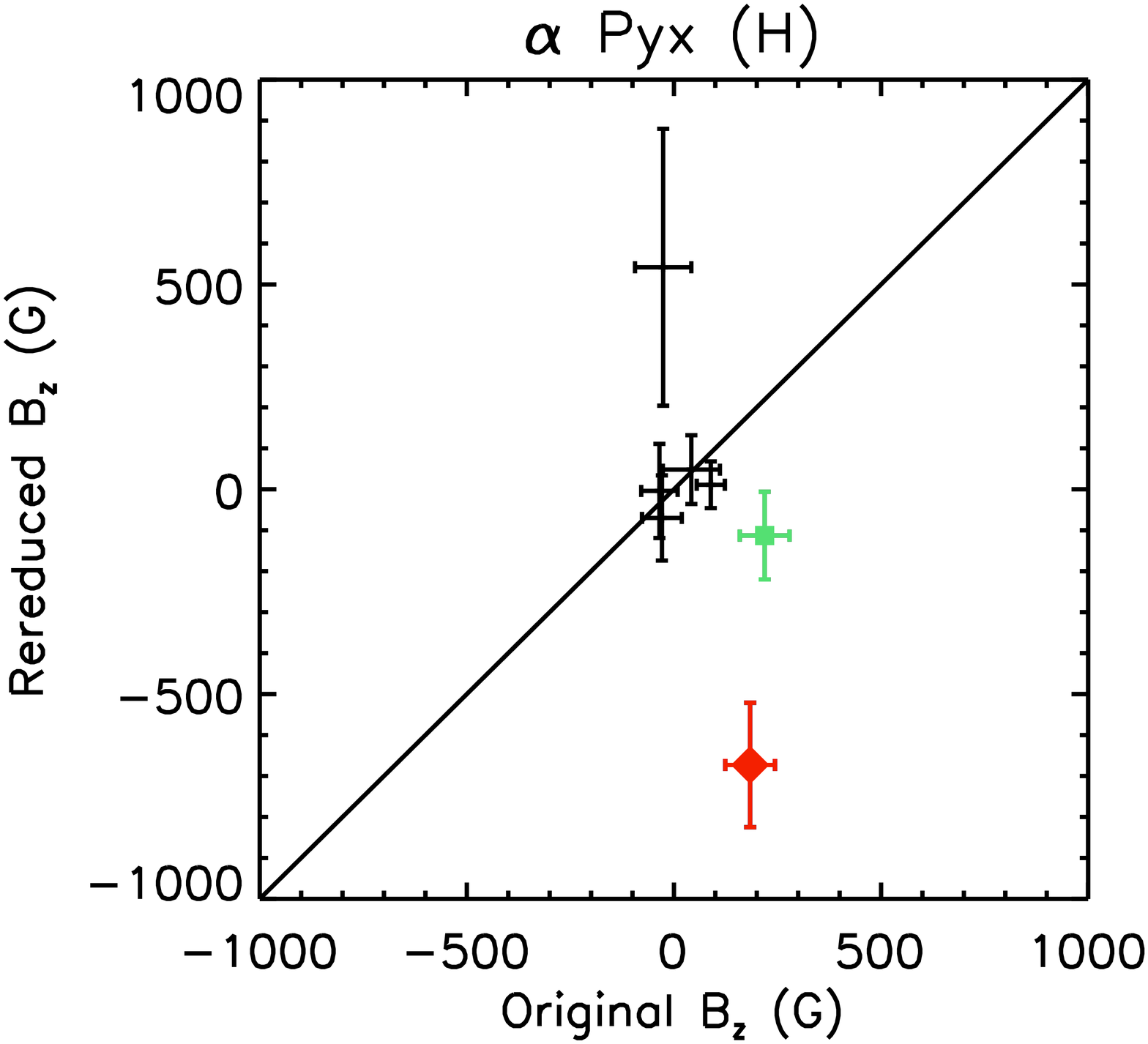}
\includegraphics[width=4.25cm]{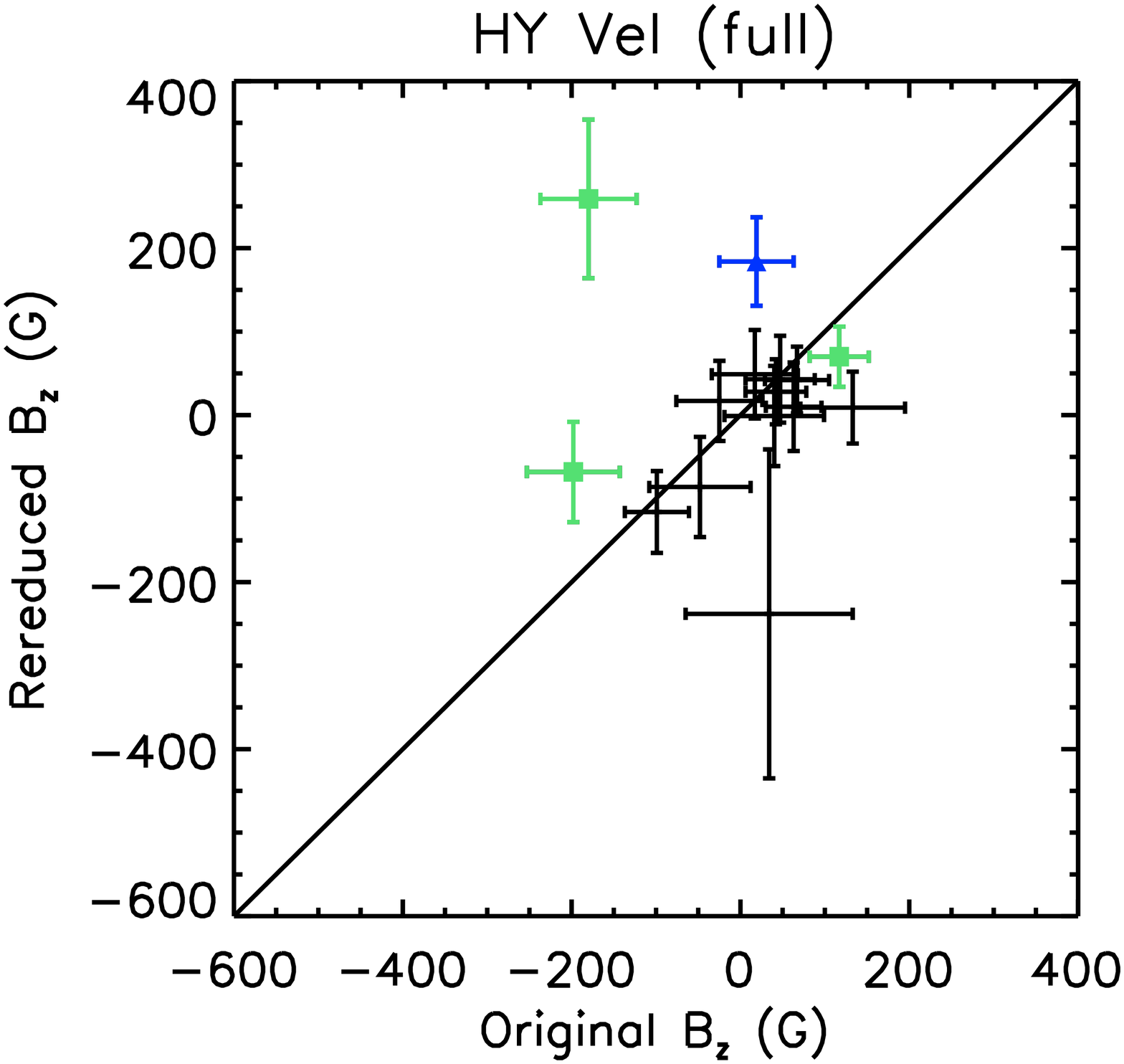}
\includegraphics[width=4.25cm]{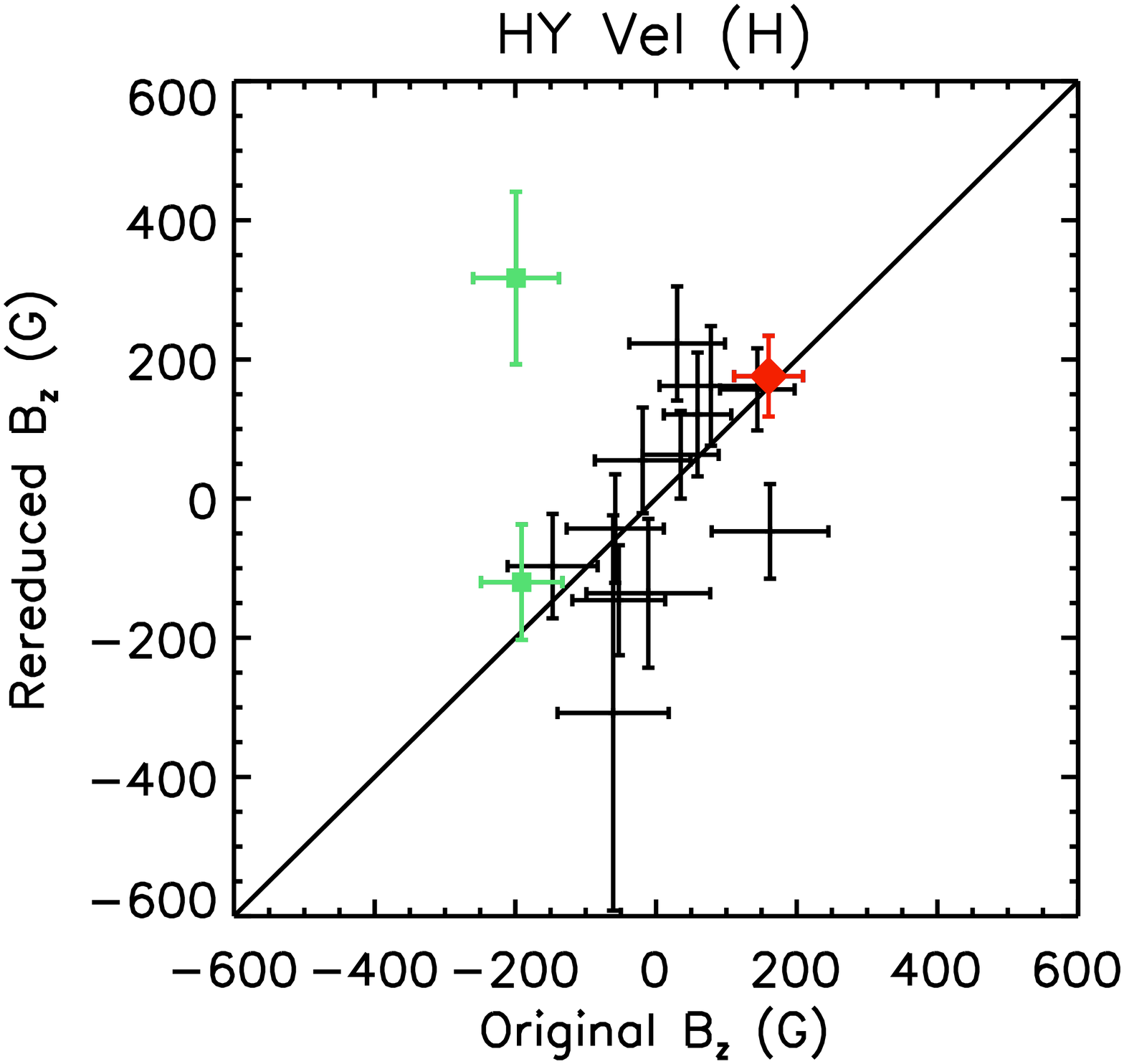}
\caption{Comparison between FORS1/2 data reductions. Left: full spectrum measurements; right: H Balmer line measurements. The straight line corresponds to $x=y$. A magnetic field is unambiguously detected in both reductions in $\epsilon$ Lupi only; in all other cases the re-reduced data are most consistent with the originally published data around $\langle B_z \rangle=0$. In the online version, 3$\sigma$ outliers are indicated in color: green for 3$\sigma$ measurements reported by H2011a, blue for 3$\sigma$ outliers in the re-reduced measurements, and red for measurements significant in both reductions. Note that some such outliers agree in significance but not polarity.}
\label{rered_compare}
\end{figure}

Examination of the published measurements of $\alpha$~Pyx, HY Vel, 33 Eri and 15 CMa suggest that signal is present in the data: 12 full spectrum measurements and 10 H line measurements are non-zero at greater than 3$\sigma$ significance. Moreover, comparison of the measurements for each star with a variety of simple models consistently favours the presence of a variable magnetic field. On the other hand, the measurements from the re-reduced FORS1/2 spectra present far fewer significant detections, and remarkably these correlate rather poorly with the detections in the original analysis. Clearly the characteristics of the new measurements differ from those previously published.

The published FORS1/2 measurements for all stars are compared to the re-reduced values in Fig. \ref{rered_compare}. Visual inspection of Fig. \ref{rered_compare} shows that while differences exist between individual $\langle B_z\rangle$ measurements, the longitudinal fields determined from the two data sets are generally in formal agreement. However, the error bars of the re-reduced data are in many cases systematically larger than those of the original published results, leading to a greater proportion of the individual measurements being formally consistent with a null result; this may be because the error bars of the re-reduced measurements are weighted with the square root of the reduced $\chi^2$. Nevertheless, the re-reduced measurements still present a number of 3$\sigma$ detections that are difficult to reconcile with Gaussian noise in the absence of a magnetic field. 

\begin{figure*}
\centering
\begin{tabular}{cccc}
\includegraphics[width=4.15cm]{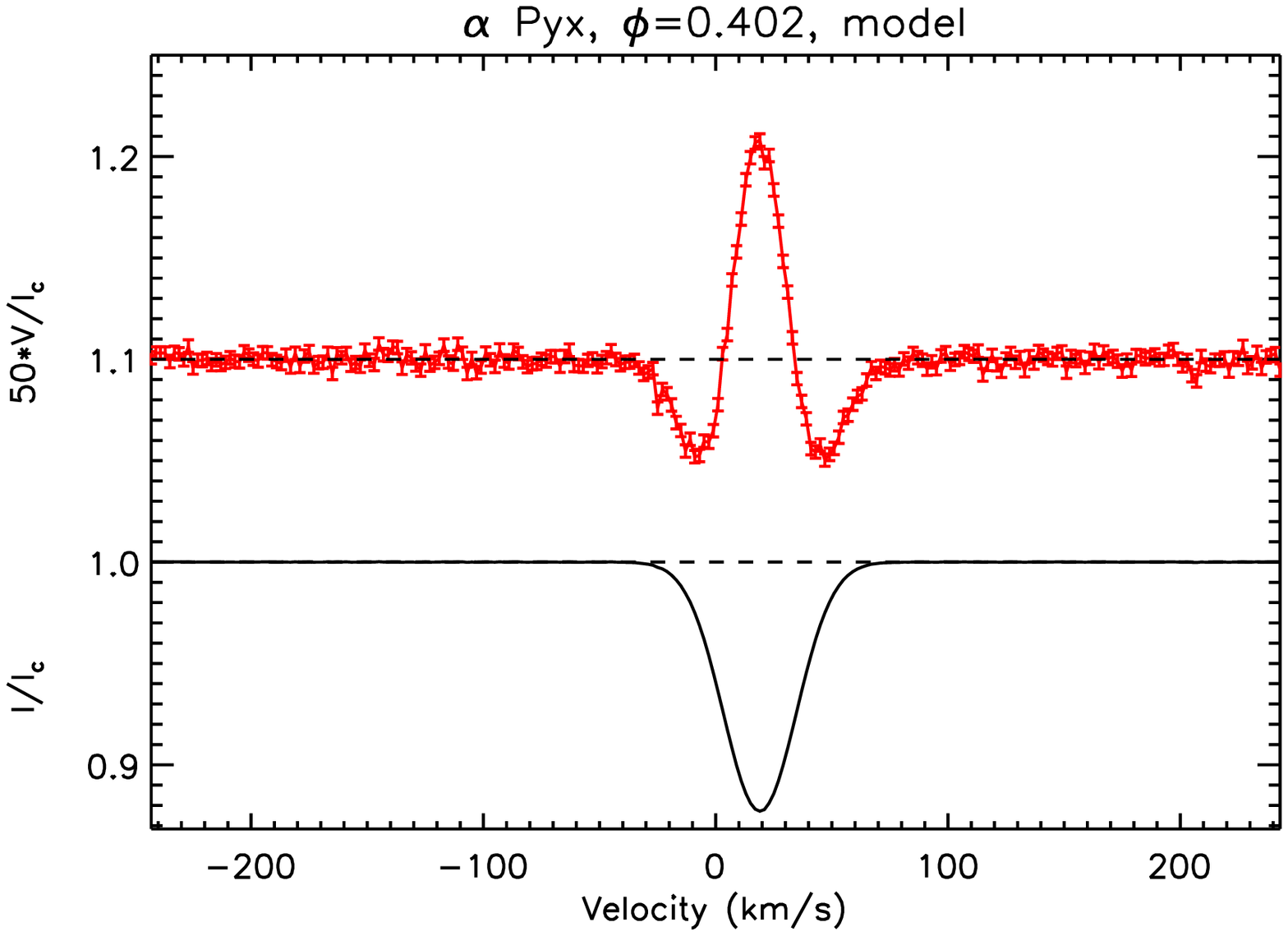} & 
\includegraphics[width=4.15cm]{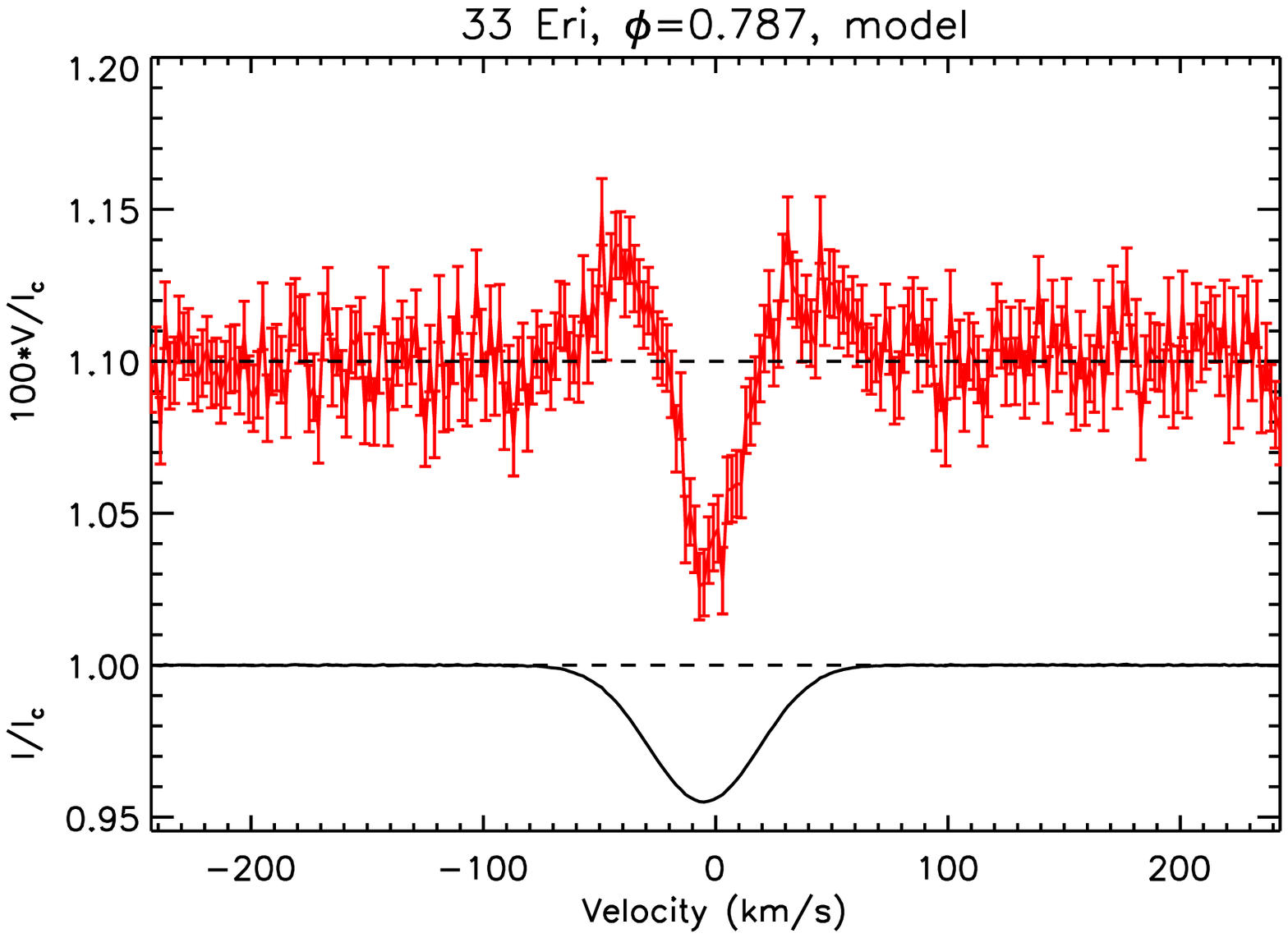} & 
\includegraphics[width=4.15cm]{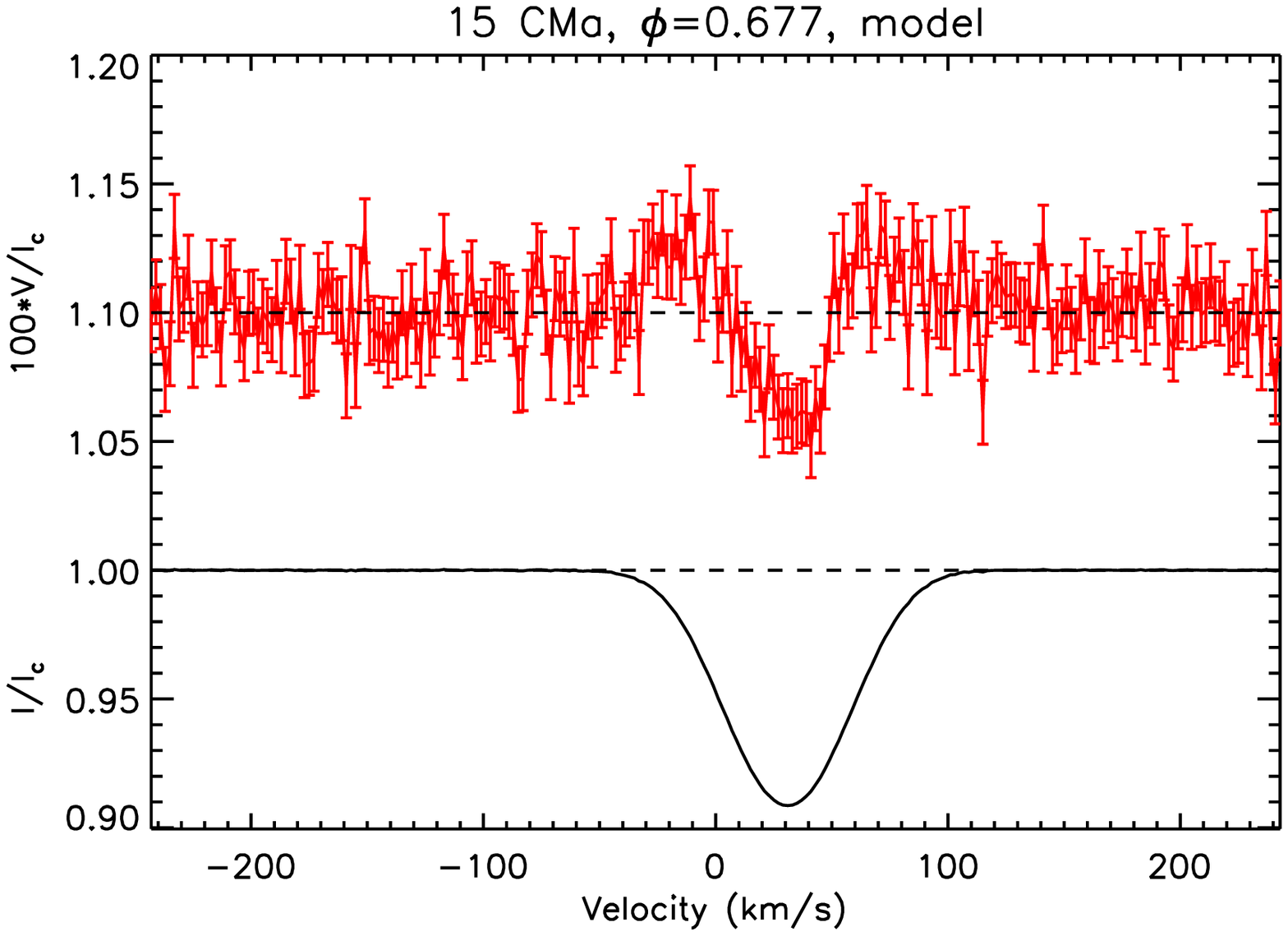} & 
\includegraphics[width=4.15cm]{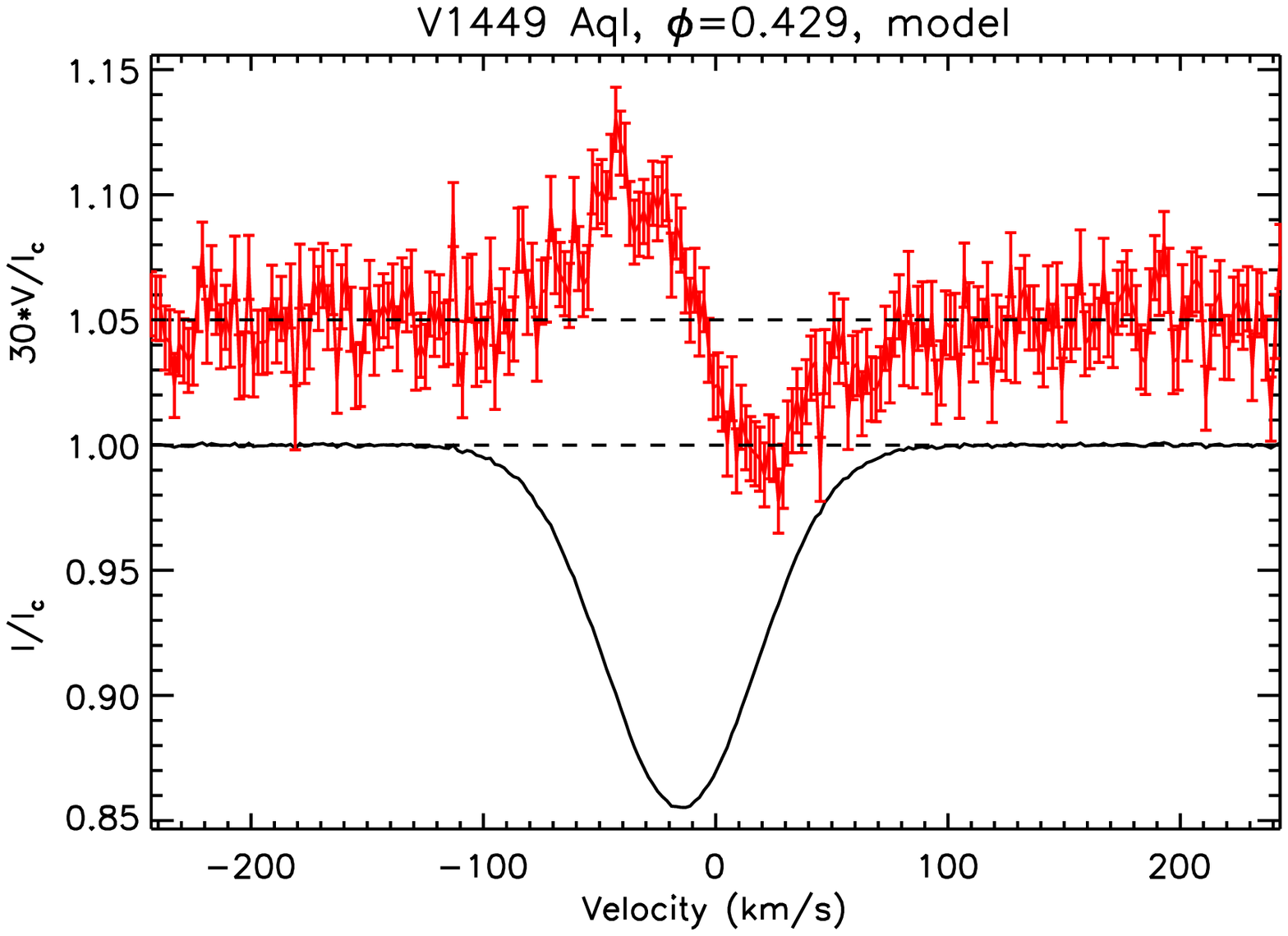} \\
\includegraphics[width=4.15cm]{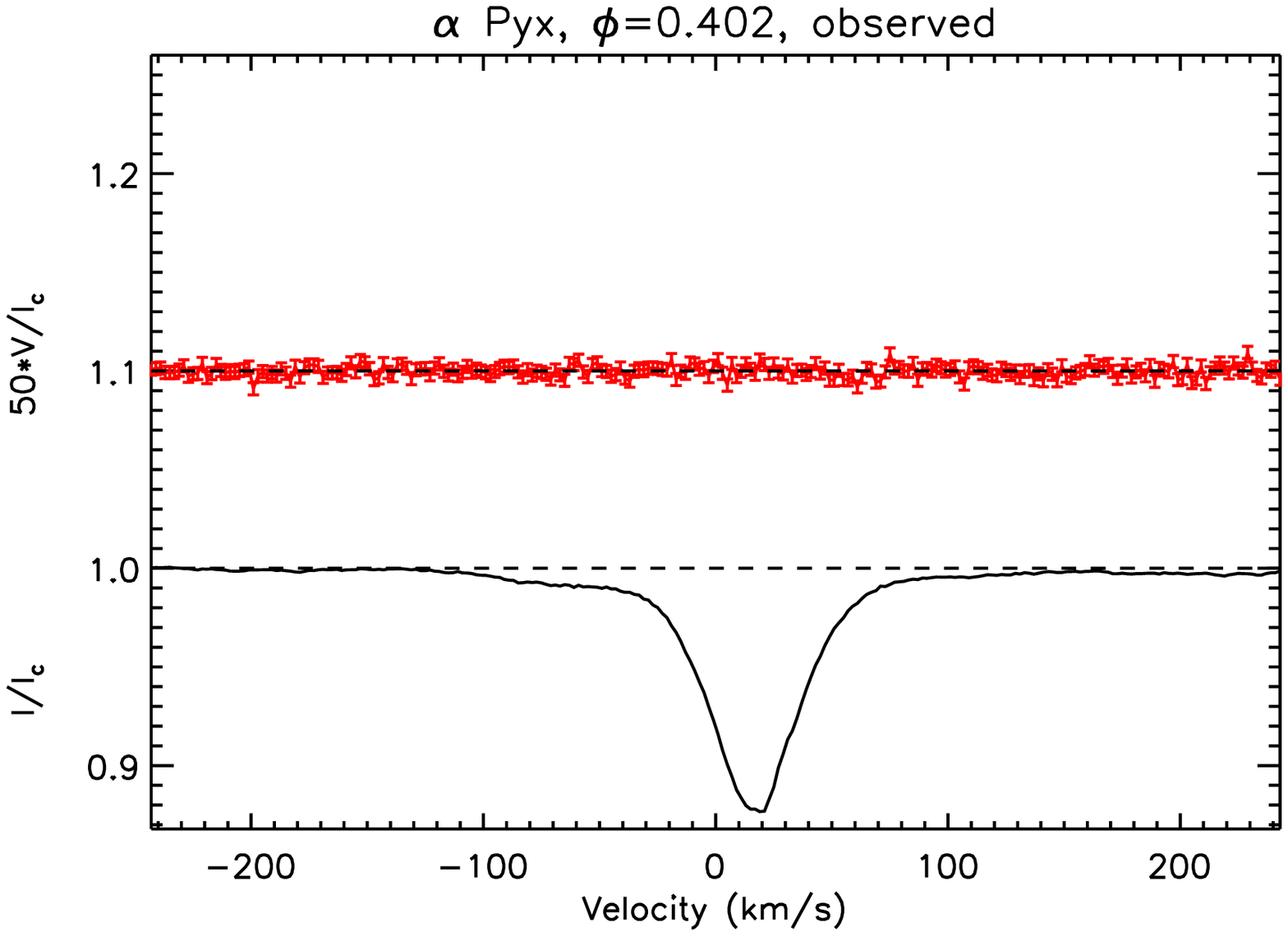} &
\includegraphics[width=4.15cm]{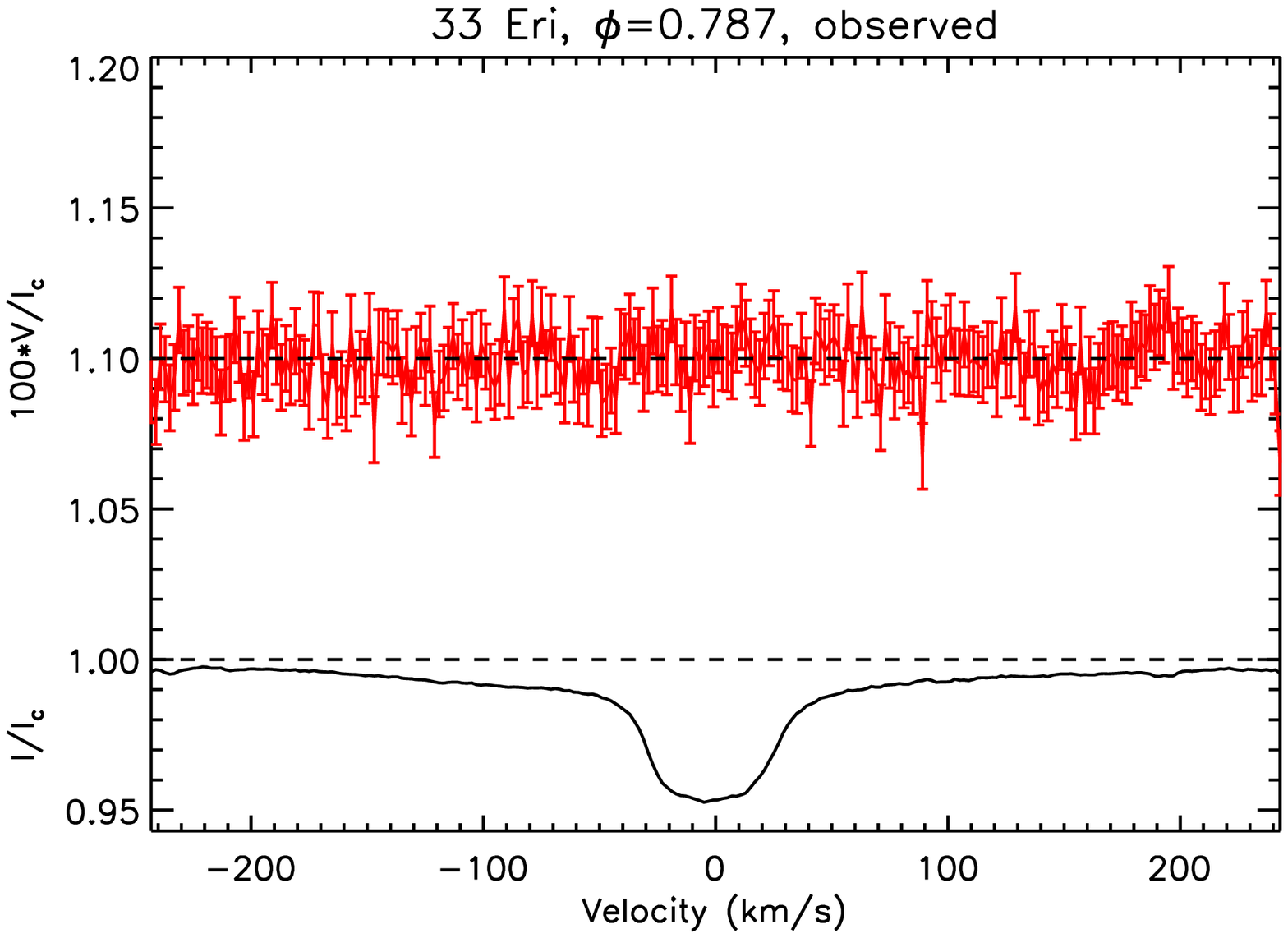} &
\includegraphics[width=4.15cm]{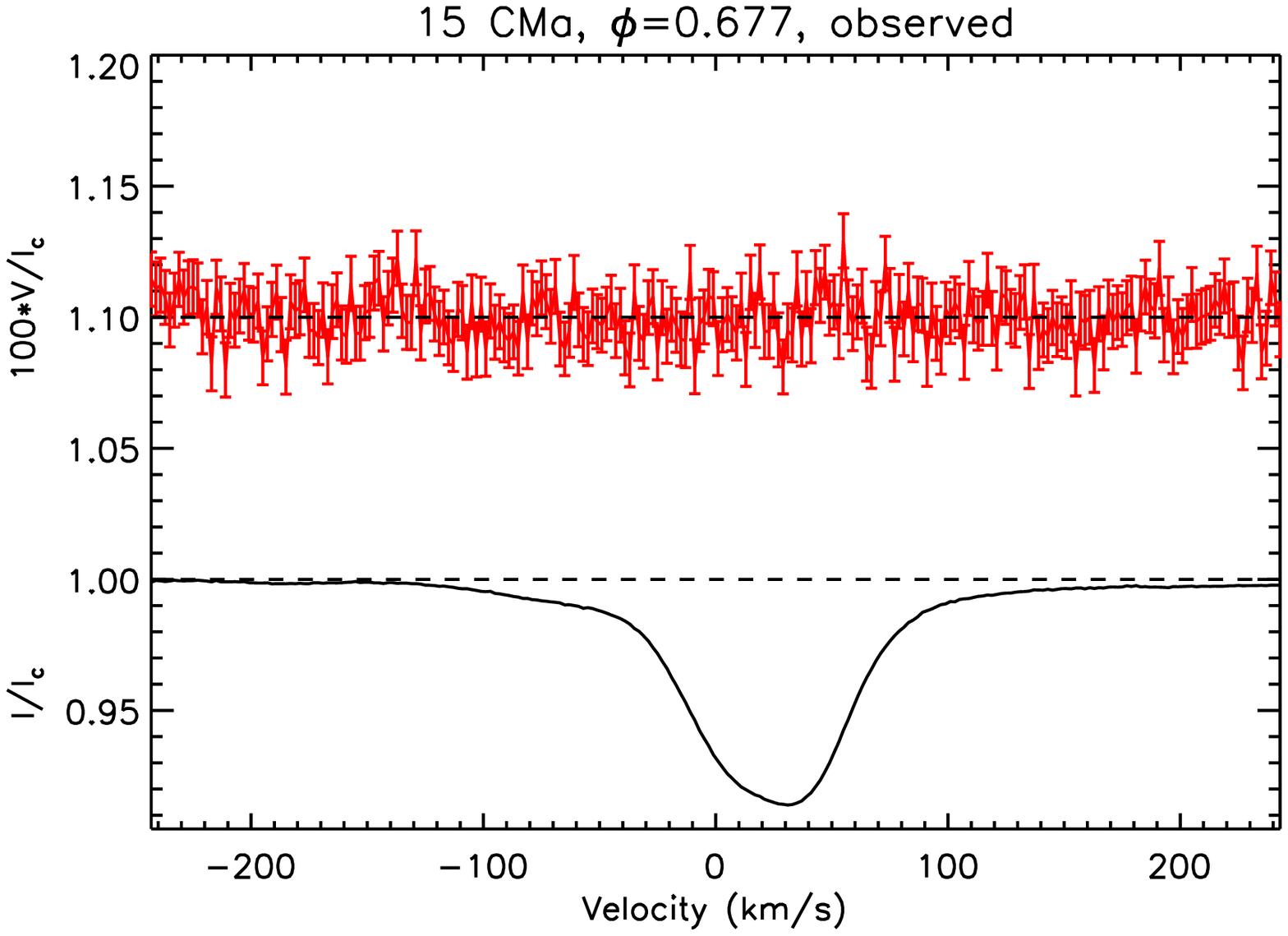} &
\includegraphics[width=4.15cm]{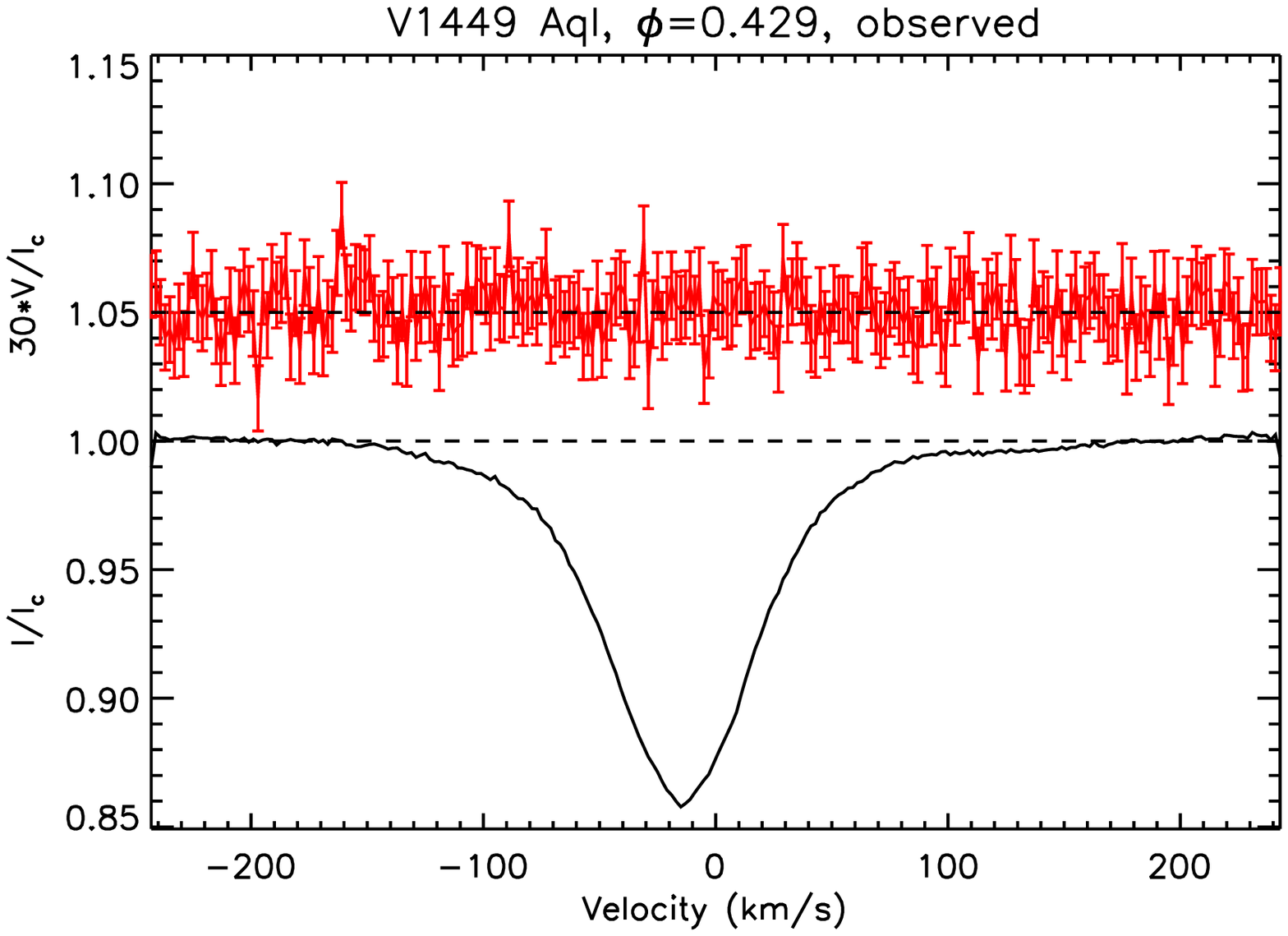} \\
\end{tabular}
\caption{Comparison of observed with synthethic LSD profiles. Synthetic LSD profiles above, corresponding observed profiles below; left--right: $\alpha$ Pyx, 33 Eri, 15 CMa, V1449 Aql. The models are created according to the parameters of H2011a and H2011b and phased according to the ephemerides therein. Turbulence has been added to improve the agreement in the wings of the intensity profile. Synthetic Gaussian noise corresponding to that of the observed profiles has been added. In all cases the Stokes $V$ signatures predicted by the proposed field configurations should have been clearly detectable at the observed phases and SNRs.}
\label{lsd_grd}
\end{figure*}

It is hard to overstate the importance of examining diagnostic nulls as well as circular polarization spectra when examining spectropolarimetric data. In the ideal case, there would be either a clear signature in Stokes \textit{V} and no signature in \textit{N} (an unambiguous detection), or no signature in Stokes \textit{V} and no signature in \textit{N} (an unambiguous non-detection.) However, it occasionally happens that there is a signature in both Stokes \textit{V} and \textit{N}, or in \textit{N} but not in Stokes \textit{V}. Field detections in the null profiles may be due to either instrument flexures and/or radial velocity shifts intrinsic to the
source. Schnerr et al. (2006) have modelled the effects of stellar pulsations on both Stokes $V$ and in the null profiles. Bagnulo et al. (2012) have argued that magnetic field measurements obtained with FORS represent a slightly more complex situation, because radial velocity shifts affect not only null profiles (and, to a lesser extent, also Stokes V profiles), but also the normalised derivative of Stokes $I$. Bagnulo et al. (2012) have concluded that while a field detection in the null profiles does not necessarily imply that a field detection in Stokes $V$ (obtained from the same dataset) is spurious, it demonstrates that photon noise is not the only contribution to the error bars.  Examining the incidence of longitudinal field detections from the diagnostic null spectra of pulsating B stars observed with FORS confirms that the error bars published by H2011a (and to a lesser degree those obtained in our re-reduction) are an underestimate of the actual error bars by a factor of 30 to 50\%.


While examination of \textit{N} is indispensable, it is not always sufficient. For instance, in the case of $\alpha$ Pyx, the FORS spectrum taken on HJD 2454109.150 yields a strong $\langle B_z\rangle$ measurement in the blue end of the spectrum but shows nothing in the Balmer lines H$\beta$--H$\delta$ or in the metal lines. When $\langle B_z\rangle$ measurements are extracted separately for H$\beta$--H$\delta$ and the higher Balmer lines, the former group yields $\langle B_z\rangle\simeq -80\pm150$ G from Stokes \textit{V} and $\sim 130\pm110$ G for \textit{N} (consistent with a null result), while the latter set of lines yields $\langle B_z\rangle\simeq-1900\pm260$ G from Stokes \textit{V} and $\sim -625\pm200$ G from \textit{N}, an apparent 7.3$\sigma$ detection (with a 3.1$\sigma$ detection in \textit{N}). This seems unlikely to happen if the circular polarization signature arises from real photospheric magnetic fields.


\section{Evaluation of Models}
\subsection{ESPaDOnS and Narval}


Recall that of the 7 pulsating B stars claimed to be magnetic by H2011a and H2011b, ephemerides are presented for 5: $\xi^1$ CMa, 33 Eri, 15 CMa, $\alpha$ Pyx and V1449 Aql. For each of these stars they derived a model of the magnetic field, including inclination of the rotation axis relative to the line of sight $i$, magnetic obliquity $\beta$ and dipolar field strength $B_{\rm d}$. 

ESPaDOnS and Narval longitudinal magnetic fields of 15 CMa, 33 Eri, $\alpha$ Pyx and V1449 Aql are uniformly consistent with the absence of any magnetic fields, in contrast to the FORS1/2 results published by H2011a and H2011b. One possible explanation of this discrepancy is suggested by an examination of Fig. \ref{fors_esp_com}. While the ESPaDOnS/Narval $\langle  B_z\rangle $ measurements are all consistent with a null longitudinal field, many of these measurements were obtained at times when, according to the proposed ephemerides and field geometry models, the longitudinal field should be close to zero. Some data points, because of the relatively low amplitudes of the model longitudinal field variations and generally larger error bars of the FORS data, are in fact in reasonable agreement with the FORS measurements obtained at those same phases. So, while some of the ESPaDOnS/Narval $\langle B_z\rangle$ measurements are strongly inconsistent with the published variations -- V1449 Aql in particular -- one can reasonably accommodate many of the remaining differences within the measurement uncertainties and possibly the uncertainties of the model parameters.

However, the longitudinal field is but one (secondary) diagnostic of the magnetic field extracted from the ESPaDOnS and Narval data. This is because a large variety of magnetic configurations can produce a mean longitudinal field component that is formally null (i.e. for which the first-order moment of Stokes $V$ is zero) at some phases. Nearly all of these configurations will generate a detectable Stokes $V$ signature in the velocity-resolved line profile. The morphology of the Stokes $V$ profile often allows these various magnetic configurations -- all of which yield the same (zero) longitudinal field -- to be distinguished. 

None of the LSD Stokes $V$ profiles for $\alpha$ Pyx, 15 CMa 33 Eri or V1449 Aql show any significant signal. To predict the Stokes $V$ profiles we would expect to observe as a consequence of the proposed field geometry models, we calculate synthetic disc-integrated Stokes $I$ and $V$ profiles. The model is essentially identical to that described by Petit et al. (2008, 2011): the local Stokes $I$ profiles are assumed to be Gaussian in shape; the local Stokes $V$ profile is computed assuming the weak-field approximation; and an explicit disc integration is performed assuming a simple limb-darkening law. Various calculated Stokes $I$ and $V$ profiles computed by the simple model have been compared to calculations obtained using the Zeeman code (Landstreet et al. 1988, Wade et al. 2001) and are found to be in acceptable agreement with the detailed polarized radiative transfer calculations. 

We proceed to compute ``synthetic observations" as follows, treating the LSD profiles as a single, mean spectral line. First, we match as closely as possible the observed and model Stokes $I$ profile. We begin with the $v\sin i$ reported by H2011a and H2011b, scale the model profile to match the depth of the observed LSD profile, then add Gaussian turbulence as necessary to reproduce as closely as possible the line wings. Then, using the mean wavelength and Land\'e factor of the real LSD profile, we compute the model Stokes $V$ profile corresponding to the appropriate phase and field geometry parameters ($i$, $\beta$, $B_{\rm d}$) of H2011a and H2011b. Finally, we resample the model profiles to the 1.8~km~s$^{-1}$ pixels of the observed profiles and add synthetic Gaussian noise corresponding to the S/N of the real LSD profile. Clearly, these fits to the LSD profiles are not meant to represent a detailed line synthesis in the complex pulsating atmosphere. However, given the success of Donati et al. (2001) in applying a similar modeling procedure to self-consistently reproduce the Stokes $V$ profiles and longitudinal field curve of $\beta$~Cep itself (which we point out appears to exhibit much larger pulsational distortion of its line profiles than the stars discussed here) we are confident that this procedure provides a reasonable first-order calculation of the amplitude and shape of the Stokes $V$ profiles predicted by the H2011a and H2011b models.

Analysing the synthetic observations using the same procedures as used for the real observations, we find longitudinal fields that are (naturally) in good agreement with the model predictions (i.e. the solid curves in Fig. \ref{fors_esp_com}). The Stokes $V$ profiles of the synthetic observations, on the other hand, are strikingly different from those observed. In every observation of 33 Eri, 15 CMa and $\alpha$~Pyx, the proposed field geometry models predict Zeeman signatures that would be easily observed and definitely detected i.e. detection probability $>99.999\%$) in our observations. These results are illustrated in Fig. \ref{lsd_grd} for LSD profiles obtained from masks including He lines; if He lines are excluded, the results are essentially the same. The observations included in this figure were selected to be those for which the longitudinal field was most compatible with the models, as determined from Fig. \ref{fors_esp_com}. The disagreement between the predicted and observed Stokes \textit{V} profiles is quite striking. Given the SNRs obtained in ESPaDOnS and Narval spectra, the models proposed by H2011a and H2011b predict definite detections of Zeeman signatures in all 15 observations of 33 Eri, 15 CMa, $\alpha$~Pyx and V1449 Aql. Therefore, unlike the longitudinal field measurements, the incompatibility of the observed Stokes $V$ profiles and those predicted by the models is totally unambiguous.

\subsection{FORS1/2 and SOFIN}

One puzzling aspect of the results presented by H2011a and H2011b is the remarkably low values of the reduced $\chi^2$ for some stars when the data are fit with a sinusoidal variation: in general a reduced $\chi^2$ significantly below 1 for a sample of this size is indicative of fitting noise or over-estimation of error bars. However, we recall that the re-reduction of the FORS1/2 spectra has yielded error bars systematically larger than those of the original reduction, suggesting that the H2011a error bars are not substantially over-estimated (and are potentially under-estimated.) 

\begin{figure}
\includegraphics[trim=10mm 10mm 10mm 10mm, clip, width=5.7cm, angle=270]{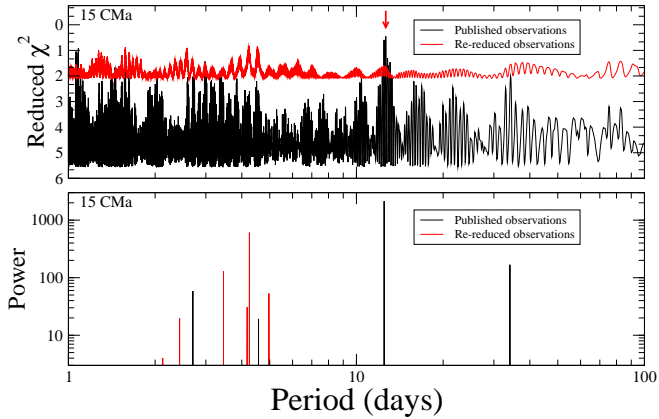}
\caption{{\bf Top:} Periodograms generated using the originally published and re-reduced FORS1/2 data. H2011 period is indicated with a red arrow. Systematically larger error bars in the re-reduction lead to a smaller reduced $\chi^2$. The periodograms share a similar variability pattern, presumably beats arising from the window function. The most significant peaks are not at the same periods. {\bf Bottom:} {\sc clean}ed periodograms. Once again, the detected periods are dissimilar between reductions.}
\label{periods}
\end{figure}


To investigate this matter further, we have constructed periodograms using both the full spectrum data and the H line data, both published and re-reduced, for all five stars for which detailed models were presented by H2011a and H2011b. We use a Lomb-Scargle type approach, computing the reduced $\chi^2$ as a function of adopted period by fitting a linear first-order sinusoid to the data. Examples of the derived periodograms are shown in Fig. \ref{periods}. The periodograms are quite complex, and in all cases numerous periods exist corresponding to fits to the data with reduced $\chi^2$s comparable to the periods reported by H2011a and H2011b. In other words, the H2011a and H2011b periods are not unique. None of these periods seems to be \textit{a priori} preferable to another. In the cases of $\alpha$ Pyx, 15 CMa and V1449 Aql, the periods identified by H2011a and H2011b are in fact those yielding the lowest reduced $\chi^2$ (although in the case of $\alpha$ Pyx there are hundreds of comparable choices). In the case of 33 Eri, the period selected does not provide the best fit; a longer period of 10.437 days yields an improved value of $\chi^2/\nu$. Ultimately we see little objective justification for selecting one amongst the various possible period solutions for these data, as there are numerous solutions providing an acceptable fit to the time series.

Because the temporal sampling of these data is diverse, the data window function is expected to be rather complex. Therefore to further explore possible periodicities in the data, we have performed an independent FFT periodogram analysis using the {\sc clean} procedure as implemented in {\sc clean-ng} (Gutierrez-Soto et al., 2009). The {\sc clean}ed periodograms typically exhibit one dominant period, however in only one case (15 CMa) is that dominant period close to that selected by H2011a and H2011b. Moreover, the different reductions of the FORS data do not yield the same periods (see Fig. \ref{periods}). Application of our period analysis procedure to synthetic time series with the same window functions and error bars but consisting only of noise show that period solutions with similar amplitudes to those obtained from the published data can be easily obtained for all stars but 15 CMa. However, we find that when the most significant published measurements of 15 CMa are removed from the analysis (i.e. those with significance larger than 2$\sigma$), the resulting periodogram continues to show a strong peak at that same period, a persistence which can only suggest that the reported period arises from a fit to the noise.


\section{Conclusions}

In this paper we have re-examined the conclusions reached by H2011a and H2011b regarding the existence of magnetic fields in 6 pulsating B-type stars, using both the measurements presented in the original papers and re-reduced archival data, as well as our own Stokes $V$ observations of these stars obtained using high-resolution ESPaDOnS and Narval spectropolarimetry. 

In one case - $\epsilon$ Lupi - Least-Squares Deconvolution applied to the high-resolution Stokes $V$ spectra confirms the presence of a magnetic field. $\epsilon$ Lupi therefore represents a newly-confirmed magnetic star for which we measure a longitudinal field value in very close agreement with the mean given by H2011a. 

For the remaining five stars in the sample studied in this paper -- 15 CMa, 33 Eri, $\alpha$ Pyx, HY Vel and V1449 Aql -- the high-resolution data do not confirm the presence of a magnetic field, although in the case of HY Vel the precision is quite poor; further observations of this star are required to reach any firm conclusions about its magnetic properties. For the four stars for which magnetic models were reported by H2011a and H2011b but for which no signal is detected in the multiple ESPaDOnS and Narval observations -- 15 CMa, 33 Eri, $\alpha$ Pyx and V1449 Aql -- the Stokes $V$ profiles predicted by the models are in clear disagreement with our observations. An examination of periodograms shows no strong reason to prefer the rotational periods derived by H2011a and H2011b over a range of other, equally probable periods, except perhaps in the case of 15 CMa, although we emphasize that for this star there is no indication of the predicted structure within the Stokes $V$ profile, even for those data which - examining merely the phased $\langle B_z\rangle$ measurements - are in apparent agreement with the FORS1/2 data. Furthermore, the periods suggested by H2011a are not robust under re-reduction of the FORS data. \textit{ Based on our examination of the published results, our re-analysis of the original archival spectra, and analysis of new high-resolution spectropolarimetry, we are unable to confirm the magnetic field detections, periods and models presented by H2011a and H2011b for 15 CMa, 33 Eri, $\alpha$ Pyx and V1449 Aql.}

Many of the inconsistencies between FORS1/2 results and those from high-resolution instruments such as ESPaDOnS or Narval can be resolved with the assumption that the formal error bars of FORS measurements have been systematically underestimated by a factor of approximately 0.5; alternatively, an increase of the FORS1/2 detection threshold from 3$\sigma$ to 6$\sigma$ would eliminate the majority of the unconfirmed detections. This conclusion is consistent with those reached by Silvester et al. (2009) and Bagnulo et al. (2012). 


H2011a point out that many of the stars discussed in this paper are both overabundant in nitrogen (e.g., Morel et al. 2008) and exhibit pulsations dominated by a non-linear radial mode. They propose that the presence of a magnetic field may explain these phenomena. Our inability to confirm the presence of magnetic fields in these stars does not support this proposal. Similarly, the large equatorial rotational velocities implied by the assumed inclination from the line of sight \textit{i} and the measured $v\sin i$ -- which are suggested by H2011a to be mostly quite high due to near pole-on inclinations for all stars but 15 CMa -- should be revisited. We ultimately conclude that no evidence exists in either the results presented by H2011a or the present work to support a correlation between photospheric magnetic fields and chemical peculiarities or pulsational phenomena of SPB and $\beta$~Cep stars: magnetism does not appear to be any more common in this class of star than in the overall population of B-type stars (approximately 10\% of early-type stars are detected as magnetic by the MiMeS survey, as summarized by Grunhut et al., 2011). This is in contrast to results for rapidly oscillating peculiar A-type (roAp) stars, for example, in which magnetic fields are consistently found to be present and related to the magnetic field via the oblique pulsator model (e.g. Kurtz et al. 1997).

\section{Acknowledgements}
J.D. Landstreet, G.A. Wade and D.A. Hanes acknowledge Discovery Grant support from the Natural Science and Engineering Research Council (NSERC) of Canada. J.H. Grunhut acknowledges financial support in the form of an Alexander Graham Bell Canada Graduate Scholarship from NSERC. This research used the facilities of the Canadian Astronomy Data Centre operated by the National Research Council of Canada with the support of the Canadian Space Agency. We thank the proprieters of the Vienna Atomic Line Database for maintaining a valuable online resource. We would like to acknowledge Bernard Leroy for writing the {\sc clean-ng} software, James Silvester for providing details of published observations of target stars, V\'eronique Petit for illuminating discussions on disc integration mechanics, and Huib Henrichs for wise advice on data presentation.

\clearpage
\clearpage
\clearpage
\clearpage
\clearpage


\begin{thebibliography}{}
\bibitem[Bagnulo(2002)]{bagn2002} Bagnulo, S., Szeifert, T., Wade, G. A. et al.,  2002, A\&A, 389, 191
\bibitem[Bagnulo(2006)]{bagn2006} Bagnulo, S., Landstreet, J. D., Mason, E. et al., 2006, A\&A, 450, 777
\bibitem[Bagnulo(2011)]{bagn2011} Bagnulo, S., Landstreet, J. D., Fossati, Kochukhov, O. 2012, A\&A, arXiv:1112.3969v1 [astro-ph.SR]
\bibitem[()]{} Bagnulo, S., Landolfi, M., Landstreet, J. D. et al., 2009, PASP, 121, 993
\bibitem[()]{} Briquet, M., Uytterhoeven, K., Morel, T., 2009, A\&A, 506, 269
\bibitem[()]{} De Cat, P. \& Aerts, C., 2002, A\&A, 393, 965
\bibitem[()]{} De Cat, P., Briquet, M., Daszy\'nska-Daszkiewicz, J. et al., 2005, A\&A, 432, 1013
\bibitem[Donati(2006)]{don2006} Donati, J.-F., Howarth, I. D., Jardine, M. M. et al., 2006, MNRAS, 370, 629
\bibitem[Donati(1997)]{don1997} Donati, J.-F., Semel, M., Carter, B. D., Rees, D. E., Collier Cameron, A., 1997, MNRAS, 291, 658
\bibitem[Donati(1992)]{don1992} Donati, J.-F., Semel, M., Rees, D. E., 1992, A\&A, 265, 669
\bibitem[()]{} Grunhut, J. H., Wade, G. A. \& the MiMeS Collaboration, 2011, proceedings of ``Four Decades of Research on Massive Stars", in press (arXiv::1109.6384)
\bibitem[()]{} Guti\'errez-Soto, J., Floquet, M., Samadi, R. et al., 2009, A\&A, 506, 133
\bibitem[Hubrig(2006)]{h2006} Hubrig, S., Briquet, M., Sch\"oller, M. et al., 2006, MNRAS, 369, 61
\bibitem[Hubrig(2009)]{h2009} Hubrig, S., Briquet, M., De Cat et al.,  2009, AN, 330, 317
\bibitem[Hubrig(2011b)]{h2011b} Hubrig, S., Ilyin, I., Briquet, M. et al., 2011b, A\&A, 531, 20
\bibitem[Hubrig(2011a)]{h2011a} Hubrig, S., Ilyin, I., Sch\"oller, M. et al., P, 2011a, ApJ, 726, 5
\bibitem[()]{} Kupka, F., Piskunov, N.E., Ryabchikova, T.A., Stempels, H.C. \& Weiss, W.W., 1999, A\&AS 138, 119
\bibitem[()]{} Kupka, F., Ryabchikova, T.A., Piskunov, N.E., Stempels, H.C. \& Weiss, W.W., 2000, Baltic Astronomy, 9, 590
\bibitem[()]{} Kurtz, D. W., Martinez, P., Tripe, P. \& Hanbury, A. G., 1997, MNRAS 289, 645
\bibitem[Landstreet(1988)]{land1988} Landstreet, J. D., 1988, ApJ, 326, 967
\bibitem[()]{} Lef\'evre, L., Marchenko, S. V., Moffat, A. F. J. \& Acker, A., 2009, A\&A, 507, 1141
\bibitem[Mathys(1989)]{m1989} Mathys, G., 1989, FCPh, 13, 143
\bibitem[()]{} Mathys, G., 1994, A\&AS, 108, 547
\bibitem[()]{} Morel, T., Butler, K., Aerts, C., Neiner, C. \& Briquet, M., 2006, A\&A, 457, 651
\bibitem[()]{} Morel, T., Hubrig, S. \& Briquet, M., 2008, A\&A, 481, 453
\bibitem[()]{} Neiner C., Alecian E., Briquet M., Floquet M., et al., 2012, A\&A, in press
\bibitem[()]{} Petit, V., Massa, D. L., Marcolino, W. L. F., Wade, G. A. \& Ignace, R., 2011, MNRAS, 412, 45
\bibitem[()]{} Petit, V., Wade, G. A., Drissen, L., Montmerle, T. \& Alecian, E., 2008, MNRAS, 387, 23
\bibitem[()]{} Piskunov N.E., Kupka F., Ryabchikova T.A., Weiss W.W. \& Jeffery C.S., 1995, A\&AS 112, 525-53
\bibitem[()]{} Rivinius, T., Wade, G. A., Townsend, R. H. D, 2011, IAUS, 272, 210
\bibitem[()]{} Ryabchikova, T.A., Piskunov, N.E., Kupka, F. \& Weiss W.W., 1997, Baltic Astronomy, 6, 244
\bibitem[()]{} Saesen, S., Briquet, M. \& Aerts, C., 2005, in ``Distant Worlds", Joint European and National Astronomy Meeting, 18, http://www.astro.ulg.ac.be/RPub/Colloques/JENAM
\bibitem[()]{} Schnerr, R. S., Verdugo, E., Henrichs, H. F., Neiner, C., 2006, A\&A 452, 969
\bibitem[()]{} Shobbrook, R. R., Handler, G., Lorenz, D. \& Mogorosi, D., 2006, MNRAS, 369, 171
\bibitem[()]{} Silvester, J., Neiner, C., Henrichs, H. F. et al., 2009, MNRAS, 398, 1505
\bibitem[()]{} Semel, M., Donati, J.-F. \& Rees, D. E., 1993, A\&A, 278, 231
\bibitem[()]{} Stankov, A. \& Handler, G., 2005, ApJS, 158, 193
\bibitem[()]{} Telting, J.H., Schrivers, C., Ilyin, I.V., Uytterhoeven, K., De Ridder, J., Aerts, C. \& Henrichs, H.F., 2006, A\&A, 452, 945
\bibitem[()]{} Wade, G. A., Bagnulo, S., Kochukhov, O. et al., 2001, A\&A, 374, 265
\bibitem[()]{} Wade, G. A., Donati, J.-F., Landstreet, J. D., Shorlin, S. L. S., 2000, MNRAS, 313, 851



\end{thebibliography}
\end{document}